\documentclass[smallextended]{svjour3}       % onecolumn (second format)
\smartqed  % flush right qed marks, e.g. at end of proof
\usepackage{graphicx}
\newcommand{\bmth}[1]{\mbox{\boldmath${#1}$}}
\usepackage{subfigure}
\usepackage{float}

\begin{document}

\title{
Close encounters of a  rotating star with    planets in  parabolic orbits   of varying inclination
and  the  formation  of  Hot Jupiters
} \subtitle{}

\titlerunning{Close encounters of a star with planets }        % if too long for running head

\author{P. B. Ivanov \and
        J. C. B. Papaloizou }


\institute{ P. B. Ivanov    \at
            Astro Space Centre, P.N. Lebedev Physical Institute, 4/32
Profsoyuznaya Street, Moscow, 117810, Russia\\
              Tel.: +7-495-3333366\\
              Fax: +7-495-333-2378 \\
              \email{pbi20@cam.ac.uk}
%  \\
      \and
       J. C. B. Papaloizou  \at
DAMTP, Centre for Mathematical Sciences, University of Cambridge,
Wilberforce Road, Cambridge CB3 0WA, United Kingdom\\
\email{ J.C.B.Papaloizou@damtp.cam.ac.uk} }

\date{Received: date / Accepted: date}

\maketitle

\begin{abstract}
%Insert your abstract here. Include keywords, PACS and mathematical
%subject classification numbers as needed.
%\keywords{Tidal interactions \and Rotating stars \and Planet formation and evolution}
In this paper we extend the theory of close encounters of a giant planet
on a parabolic orbit with a central star developed in our previous
work (Ivanov \& Papaloizou 2004, 2007)
to include the effects of tides induced on the central star.
Stellar rotation and orbits with arbitrary inclination to the stellar rotation axis
are considered. We obtain results both from an analytic  treatment
that incorporates first order corrections to normal mode frequencies arising from
stellar rotation and numerical treatments that are in satisfactory agreement over
the parameter space of interest. These results are applied to the initial phase of the
tidal circularisation problem. We find that both tides induced in the star and planet
can lead to a significant decrease of the orbital semi-major axis
for orbits having periastron distances smaller than 5-6 stellar radii with
tides in the star being much stronger for retrograde orbits compared to prograde orbits. Assuming that 
combined action of dynamic and quasi-static tides could lead to the total 
circularisation of orbits this corresponds to observed periods up to
4-5 days.

We use the simple Skumanich law to characterise the rotational history
of the star assuming that the star has its rotational period equal to
one month at the age of 5Gyr. 
The strength of tidal interactions is characterised by
circularisation time scale, $t_{ev}$, which is defined as a typical
time scale of evolution of the planet's semi-major axis due to tides. 
This is considered as a function of orbital period $P_{obs}$, which
the planet obtains after the process of tidal circularisation has been completed.  

We find that the ratio of the initial circularisation time scales 
corresponding to prograde and retrograde orbits, respectively, is of
 order 1.5-2 for a planet of one Jupiter mass having
$P_{obs}\sim $ four days. The ratio grows with the  mass of the planet, being
of order  five for a five Jupiter mass planet with the same $P_{orb}$.  
Note, however, this result might change for more realistic stellar rotation
histories. Thus, the effect of stellar rotation may provide a bias in
the formation of planetary systems having planets on close
orbits around their host stars, as a consequence of  planet-planet scattering,
which favours systems with retrograde orbits.  
The results reported in the paper may also be applied to the problem
of tidal capture of stars in young stellar clusters.

\end{abstract}


\section{Introduction}
\label{intro}

One  explanation for the existence of  closely orbiting extrasolar giant planets is through the mechanism of orbital migration.
induced through the gravitational interaction  with the protostellar disc  (e.g. Lin \& Papaloizou 1986).
Notably this mechanism  is expected to produce planets in circular orbits that are coplanar with the the stellar equator.
However, the presence of high orbital eccentricities amongst extra
solar planets is indicative  of a strong orbital relaxation or 
scattering process.  This led Weidenschilling \& Mazari (1996) and Rasio \& Ford (1996) to consider planet-planet scattering
of planets initially on nearby circular orbits that resulted in one planet being scattered onto a highly eccentric orbit that had a close
approach to the central star.  Subsequent tidal interaction with the central star was then postulated to lead to the 
formation of  a Hot Jupiter.   Papaloizou \& Terquem (2001) investigated the strong dynamical relaxation
of a planetary system with 5 to 100 planets
starting on closely packed  circular orbits.   This type of relaxation also readily
produced planets   in a highly eccentric orbits with  close enough approaches to the central star for stellar
tides to lead to orbital circularisation. The evolution of the planet
orbit due to combined action of planet-planet scatterings and tides
has also been considered (e.g. Nagasawa et al 2008, Marchi et al 2009).

More recently measurements of the inclination of the orbital planes of transiting protoplanets
through the Rossiter-McLaughlin effect has led to the discovery of a significant  number of close orbiters 
on highly inclined or even retrograde orbits (see eg. Winn et al. 2010, Hebrard 2011,  and references therein).
This indicates that a planet-planet interaction  that can lead to high inclination orbits with high eccentricity
that are subsequently circularised  has been active in 
some cases at least.  Thus tidal circularisation of orbits with arbitrary inclination with respect to the central star need to be considered.
Tides induced in the planet are expected to be independent of orbital inclination because of the rapid attainment 
of pseudo synchronisation (see Ivanov \& Papaloizou 2007). However,  this is not the case for tides induced on the star.
For example in a frame rotating with the star, tidal forcing from a retrograde orbit will appear to have a higher forcing frequency  than 
tidal forcing from a prograde orbit. Accordingly the tidal response
will differ. This  leads to a larger transfer of energy from the planet
orbit to the star during a single periastron flyby (see also Lai 1997)\footnote{As far as we are aware, Lai was the first to
 point out that energy transfer during a flyby on a retrograde orbit is enhanced compared to the prograde case. 
In this work the  orbital plane  was taken to  coincide with  the equatorial  plane.  Here we  allow for the general 
case of arbitrary mutual 
orientations of these planes.} 
This type of phenomenon may modulate
the distribution of observed Hot Jupiters on inclined circular  orbits that may have been circularised with  tides
acting on the central star playing a significant role.

It is the purpose of this paper to study the tidal interactions of planets on highly eccentric orbits,
with orbital planes arbitrarily inclined to the stellar 
equatorial plane, taking account of tides induced on the central star and estimating their contribution to the
circularisation process. In making these studies we follow the approach of previous papers
(Ivanov \& Papaloizou 2004, 2007; Papaloizou \& Ivanov 2010)  and approximate  the initial stages of orbital 
circularisation as being through  a sequence of encounters of the
planet  on a parabolic orbit with the star. We find typical time
scales, $t_{ev}$, of evolution of the planet's semi-major axis $a$ due to dynamic tide exerted in the planet and in
the star, as a function of the period $P_{obs}$, which the planet has
after the circularisation has been completed and the planet has
settled on a low eccentricity orbit. This period can be obtained from observations.

It is important to note that the  aspects of dynamical 
 tides considered in this paper can only efficiently decrease the planet's semi-major axis
 when either the so-called stochastic instability (see e.g.  Kochanek (1992), Kosovichev $\&$ Novikov (1992), Mardling (1995)a,b,  
Ivanov $\&$ Papaloizou (2004), (2007))  operates in the dynamical system consisting of the orbit and tidally 
perturbed normal modes of the planet and the star,  or when the normal modes are able to dissipate the energy transferred
from the orbit during a time of order of or smaller than one orbital period. As we discuss below both processes can be
effective when  the semi-major axis of the orbit is sufficiently large, of the order of several $au.$ Therefore, in general, 
the time $t_{ev}$ characterises only the orbital evolution at such scales.  At smaller scales,  nonlinear effects not 
considered here, such as wave breaking, or  frictional 
processes acting on  quasi-static tides can further
decrease $a$ thus completing the process of circularisation and
leading to the formation of close-in giant planets with small $P_{obs}$ and eccentricities (e.g. Barker 
$\&$ Ogilvie 2009 and references therein). 
Provided that typical time scale of orbital circularisation due to these processes
is smaller than $t_{ev},$  this characteristic time scale may be considered as giving an overall time 
for the planet to circularise from an initially large semi-major axis $a.$ 
In the opposite limit  the overall time scale is obviously
given by the  time scale associated with the later evolution.

In order to evaluate $t_{ev}$ we have to know the dependence of the
stellar angular velocity, $\Omega $, on time $t$. In what
follows we use the simplest possible form of such a dependence
assuming the so-called the Skumanich (1972) law for stellar rotation 
$\Omega \propto t^{-1/2}$, where the coefficient of proportionally is
chosen in such a way that the star has a period of rotation of  one month at an  age
of $5Gyr$ corresponding to that of the Sun. 

We find $t_{ev}$ to be different for prograde and retrograde
orbits, with the difference increasing with planet mass. The
ratio of the evolution time scale corresponding to a prograde orbit
to that corresponding  to a retrograde orbit is of  order  $1.5-2$ 
for a one Jupiter mass planet with a typical $P_{obs}\sim 4d$, and it
is of  order  $5$ for a five Jupiter mass planet with  the same
period. This difference gets more pronounced in a situation where tides 
exerted in the planet are not taken into account. Since conditions
determining the long term effectiveness of the transfer of energy from the orbit to 
modes of pulsation  can be different  for the planet and  star, 
such a situation may indeed occur. 
The results reported in this paper may be directly 
used in  numerical experiments involving  planet-planet
scattering which  also include tidal interactions between planets and the host star.

The plan of  the paper is as follows.
We develop an analytic theory for calculating the tidal response of a slowly rotating star
due to close the passage  of  a planetary perturber in a parabolic orbit in section \ref{sec:1}.  The orbital plane 
is allowed to have an  arbitrary inclination
to the stellar equatorial plane and
corrections to stellar normal mode frequencies correct to first order in the stellar
angular velocity are taken into account.
 We give expressions for the energy transferred to the  star in
section \ref{energytr} and the analogous expressions for the transferred angular momentum in Appendix B. 
In section  \ref{Numerical} we describe numerical solutions of the linear tidal encounter problem.
Following our previous work (Papaloizou \& Ivanov 2010)
we  take account of Coriolis forces, but not centrifugal forces
so that  the tidally undisturbed  model was spherical.
However, apart from this, the stellar angular velocity may take any value.  
 We perform  calculations  for a polytrope of index $n=3$ and unlike in our previous work
 we include a treatment of the self-gravity of the star.
 
Results for the energy and angular momentum transferred
to the star as a consequence of  prograde encounters, retrograde encounters and encounters with orbital plane
perpendicular to the equatorial plane  are then given in
section \ref{Numres}.  We then go on to  compare numerical and analytic  results in
section \ref{Numanalcomp}.  In both approaches the tidal transfers are significantly larger for
retrograde encounters as compared to prograde encounters. 
We use our results to estimate 
the initial time scale for tidal circularisation  
for an orbit with an initial semi-major axis of $10au$  that circularises in a short period orbit in section \ref{circtime}
for planets of one and five Jupiter masses.
We give estimates  with and without contributions arising from tides induced 
in the planet obtained from our previous work (eg. Ivanov \& Papaloizou 2007).
The central star is taken to be a solar mass star with angular velocity
determined by the Skumanich (1972) law.
We find that tides induced in the star and planet may separately account for the
initial circularisation of orbits down to periods of a few days.
Finally in section \ref{Discuss} we summarize and our discuss our results
and their application to the tidal circularisation of prograde and retrograde Hot Jupiters.

\section{The parabolic encounter of a planet with a rotating star: Analytic development for slow rotation
and orbits of arbitrary inclination}
\label{sec:1}
% Text with citations \cite{RefB} and \cite{RefJ}.
We develop the theory of the tidal interaction
of a planet on a parabolic orbit with a central star.
When the star is  not rotating the tidal interaction excites $p,$ $f$ and $g$ modes.
A slow rotation approximation is then adopted in which the first order corrections
to the mode frequencies due to rotation is taken into account in calculating the tidal response
for an orbit with arbitrary inclination.
\subsection{Coordinate systems}
 We calculate the energy  transfer occurring 
as a result of the action of dynamic tides induced in
a rotating star by a planet moving in a highly eccentric
orbit  in a plane that is  inclined with respect to the
stellar equatorial plane.  The eccentricity is taken to be very  close to unity
so that to a good approximation the planet may be viewed as
undergoing a sequence of parabolic encounters with the central star.  

 For such a configuration, there are two natural
Cartesian coordinate systems with origin at the  centre of the star. 
The first,
$(X^{\prime},Y^{\prime},Z^{\prime}),$ is such that
the $(X^{\prime},Y^{\prime})$ plane  coincides with the orbital plane. 
The second,  $(X,Y,Z),$ is such that
the $(X,Y)$ plane coincides with 
the equatorial plane of the star.  It is assumed that
the   the line of nodes which coincides with the Y axis
makes an  angle $\gamma $
with the $Y^{\prime}$ axis and lies in the interval 
$[0,2\pi].$  The direction of orbital motion is assumed to be
correspond to a rotation about the $Z'$ axis  when viewed in the 
$(X^{\prime},Y^{\prime})$ plane with  the
apsidal line coinciding with the  $X^{\prime}$ axis and pericentre 
such that  $X^{\prime} > 0$. Thus the  $(X^{\prime},Y^{\prime})$ plane coincides with
the orbital plane with the orbit being symmetric with respect to the 
$X^{\prime}$ axis.  
The  $(X,Y)$ plane  coincides with the
star's equatorial plane  with the  $Z$ axis directed along the
star's rotation axis. The line of intersection of the 
$(X^{\prime},Y^{\prime})$ plane  and the $(X,Y)$
plane  is chosen to define  the  $Y$ axis.

The angle $\beta $ between the $Z^{\prime})$ and $Z$ axes is
called the inclination angle. Its values lie in the range
$[0,\pi]$. Clearly, $\beta=\pi $ corresponds to a situation when
the planet's orbit lies in the equatorial plane of the star but
the direction of orbital motion is opposite to the direction of
stellar rotation. It is easy to see that the mutual orientation of
the coordinate systems  defined in this way  corresponds to the
angles $\beta $ and $\gamma $ being the usual Euler 
angles that  are  adopted to define the relative orientation
of  two coordinate systems\footnote{Note that the definition of the
Euler angles correspond to the so-called $zyz$ convention. Another way 
of defining the angles would be to perform the first rotation about
the $X$ axis and the second one about the $Z$ axis. The so-defined
Euler angles $\beta^{'}$ and $\gamma^{'}$ correspond to the frequently used $zxz$
convention. They are related to the angles used in this paper as
$\beta^{'}=\beta $ and $\gamma^{'}=\gamma -\pi/2 $.}.
Additionally, we make use of  two spherical  polar coordinate systems with 
spherical angles $(\theta^{\prime}, \phi^{\prime})$ and $(\theta,
\phi)$  respectively. The  pairs of angles  $(\theta^{\prime}, \phi^{\prime})$
and  $(\theta, \phi)$
are defined  in the usual way, but  with respect to the coordinate systems
$(X^{\prime},Y^{\prime},Z^{\prime})$ and  $(X,Y,Z),$ respectively.

\subsection{The tidal potential}

The  quantity  that determines  the energy and angular
momentum transferred between planet and star  is the tidal potential, $U$. 
When the
  $(X^{\prime},Y^{\prime},Z^{\prime})$  system is used
the tidal potential has the standard form (see e.g. Press $\&$
Teukolsky 1977, hereafter PT)
\begin{equation}
U={GMr^2\over R^3(t)}\sum^{'}_{m}
W_{m}e^{-im\Phi(t)}Y_{2,m}(\theta^{\prime},\phi^{\prime}),
\label{e1}
\end{equation}
where $G$ is the gravitational constant,  $M$ is the planet's mass and
$r$ is the spherical polar  radial coordinate. $R(t)$ and $\Phi (t)$ are
the distance between the planet and the star and the angle between
the position vector of the planet and the $X^{'}$ axis
measured in the orbital plane  
respectively. The  time
$t=0$ corresponds to a  passage through  periastron by the planet.
 The
prime above the summation sign implies hereafter that the summation is
performed over $m=0$, $\pm 2$ only. For $|m|=2$,
$W_{m}=\sqrt{{3\pi \over 10}}$ and $W_{0}=-\sqrt {\pi \over 5}$,
with $Y_{l,m}$  being the usual spherical functions.

When we consider the tidal potential in the 
$(X^{\prime},Y^{\prime},Z^{\prime})$ coordinate system
 with orientation determined by  the stellar rotation axis, its form is modified. In
order to find the new form of $U$ we can use the transformation
law of spherical functions under a rotation  of the coordinate
system. This may be written
\begin{equation}
Y_{2,m}(\theta^{\prime},\phi^{\prime})=e^{-im\gamma
}\sum_{n}d_{n,m}^{(2)}Y_{2,n}(\theta,\phi), \label{e2}
\end{equation}
where $d_{n,m}^{(2)}$ are the so-called Wigner $d$ functions, see
e.g. Varshalovich et al (1989).  Here the  summation is performed over all
integers, $n,$  such that $n\in [-2,2].$ Note that the $d$ functions depend only on the
inclination angle, $\beta,  $ and satisfy the condition  $d_{n,-m}^{(2)}=
(-1)^{n-m}d_{-n,m}^{(2)}$. Their explicit forms are given  in
Appendix A.

Substituting equation (\ref{e2}) in (\ref{e1}) we get
\begin{equation}
U=r^2\sum_{n} A_{n}Y_{2,n}(\theta,\phi), \label{e3}
\end{equation}
where
\begin{equation}
A_{n}={GM\over R(t)^{3}}\sum^{\prime}_{m}W_{m}e^{-im(\gamma +\Phi
(t))}d^{(2)}_{n,m}.\label{e4}
\end{equation}
Note that from the requirement that $U$ is a real quantity and the
properties of the spherical functions that
$Y^{*}_{2,n}=(-1)^{n}Y_{2,-n}$, where $(*)$ stands for the complex
conjugate it follows that
\begin{equation}
A^{*}_{n}=(-1)^{n}A_{-n}. \label{e5}
\end{equation}

In order to evaluate the effects of a single encounter
of the planet with the star, 
we introduce the Fourier transform in time of any quantity of
interest, say $F(t)$, $\tilde F(\sigma)$, by the relations
\begin{equation}
F(t)=\int^{+\infty}_{-\infty}d\sigma \exp(-i\sigma t)\tilde
F(\sigma ),\quad \tilde F(\sigma )={1\over
2\pi}\int^{+\infty}_{-\infty}dt\exp(i\sigma t)F(t).
\label{e6}
\end{equation}
From equations (\ref{e5}) and (\ref{e6}) it follows that the
Fourier transforms of the amplitudes $A_{n}$ obey the symmetry
relations
\begin{equation}
\tilde A^{*}_{n}(-\sigma ) =(-1)^{n}\tilde A_{-n}(\sigma).
\label{e7}
\end{equation}

As follows from equations (\ref{e1}) and (\ref{e3}) the time
dependence of the tidal potential is determined by the quantities
$I_{m}=(R_{min}/R(t))^3e^{-im\Phi (t)}$, where $R_{min}$ is
the periastron distance. Assuming that the orbit is formally
parabolic, $R(t)$ and $\Phi(t)$ can be obtained  from
the parametric representations 
\begin{equation}
\frac{R\hspace{4mm}  }{R_{min}} = 1+x^2\label{paroR}
\end{equation}
\begin{equation}
\Phi=2\tan^{-1}x\label{paroPh}
\end{equation}
\begin{equation}
\Omega_{orb}t=\sqrt{2}\left(x+\frac{x^3}{3}\right) \hspace{4mm} -\infty < t < \infty\label{parot}
\end{equation}
Using these the Fourier transforms of these quantities, $\tilde
I_{m}$, can be readily obtained in the form
\begin{equation}
\tilde I_{m}={\sqrt 2\over \pi \Omega_{orb}}I_{2,-m}(\sigma
\Omega_{orb}^{-1}), \label{e8}
\end{equation}
where
\begin{equation}
\Omega_{orb}=\sqrt {(1+q) {GM_{*}\over R^3_{min}}} \label{e9}
\end{equation}
it a characteristic angular frequency of periastron passage,
$M_{*}$ is the mass of the star, $q=M/M_{*}\ll 1$ is the mass
ratio, and
\begin{equation}
I_{2,m}(y)=\int^{\infty}_{0}{dx\over (1+x^2)^{2}}\cos (\sqrt
2y(x+x^3/3)+2m\arctan x), \label{e10}
\end{equation}
see e.g. PT\footnote{Note a misprint in Ivanov $\&$ Papaloizou
(2007). In their equation (61) equivalent to (\ref{e10}) there should
be $I_{2,m}$ on the left hand side instead of $I_{2,-m}$ as in the
paper.}. Using equations (\ref{e4}) and the definition of
$\tilde I_m$ we obtain as explicit expression for $\tilde
A_n(\sigma)$,
\begin{equation}
\tilde A_n(\sigma)={\sqrt 2\over \pi \Omega_{orb}}{GM\over
R_{min}^{3}}a_n,\quad a_n=\sum^{\prime}_{m}W_{m}e^{-im
\gamma}d^{(2)}_{n,m} I_{2,-m}.\label{e10a}
\end{equation}

\subsection{Equations of motion}

As has been discussed in our previous papers (Ivanov $\&$
Papaloizou 2004, Papaloizou$\&$Ivanov 2005, Ivanov$\&$Papaloizou
2007, Ivanov $\&$ Papaloizou 2010, Papaloizou$\&$Ivanov 2010) for
the case of coplanar encounters a perturber moving on eccentric
orbit excites normal modes of a rotating body after the periastron
passage. We must, accordingly, solve equations of motion of
stellar perturbations associated with the modes under the forcing
determined by the tidal potential in order to find the transfers
of energy and angular momentum from the orbit to the modes. As we
discuss above in this paper we would like to relax the assumption
of coplanarity of the orbital and equatorial planes treating,
however, the effect of stellar rotation in a simplest possible way
and assuming that the angular frequency of the star, $\Omega $, is
much smaller that angular frequencies of the normal modes,
$\omega_{nm}$, mainly excited in course of the tidal encounter,
$\Omega \ll \omega_{nm}$. In this case the normal modes
frequencies are close to the ones of non-rotating star,
$\omega_a$, where the index $a$ labels different types of normal
modes of a non-rotating star (e.g. p-modes, g-modes and the
fundamental mode). In the linear approximation in small parameter
$\Omega/\omega_a$ only the modes frequencies acquire corrections
of order of $\Omega $ due to rotation while other quantities
remain the same as in the non-rotating case. In the same
approximation the total energy and angular momentum transfers can
be represented as a direct sum of partial contributions
corresponding to modes of different type, and, therefore, we
derive below these expressions for a mode of one particular type,
omitting, for simplicity, the index $a$ below, thus e.g.
$\omega\equiv \omega_a$, etc..

The stellar perturbations can be described in terms of the
Lagrangian displacement vector, ${\mbox{\boldmath $\xi$}}(t,{\bf r})$. Since the
tidal potential is represented as a sum of the spherical harmonics
$Y_{l,n}$ having $l=2$ and $-2 \le n \le 2$ (see equation
(\ref{e3})) in order to solve the response problem for a mode
having in the non-rotating limit a frequency $\omega $ and $l=2$
we should consider the vector ${\mbox{\boldmath $\xi$}}(t,{\bf r})$  as a linear
combination of the eigen vectors with the same values of $n$,
\begin{equation}
{\mbox{\boldmath $\xi$}}=\sum_{n}b_n(t){\mbox{\boldmath $\xi$}}_n({\bf r}), \label{e11}
\end{equation}
where
\begin{equation}
{\mbox{\boldmath $\xi$}}_n({\bf r})=\xi_{r}(r)Y_{2,n}{\bf e}_{r}+\xi_{s}(r)(r\nabla
Y_{2,n}), \label{e12}
\end{equation}
${\bf e}_{r}$ is unit vector in the radial direction, the
functions $\xi_{r}$ and $\xi_{s}$ can be found by solving the
eigenvalue problem for a non-rotating star, see e.g. PT, Lee $\&$
Ostriker 1986. It is assumed hereafter that the eigen-vectors of
the star are normalised by the condition (PT)
\begin{equation}
\int d^3x \rho ({\mbox{\boldmath $\xi$}}_n^{*}\cdot {\mbox{\boldmath $\xi$}}_n)= \int_0^{R_*} r^2dr \rho
(\xi_r^2+6\xi_s^2)=1,   \label{e12a}
\end{equation}
where integration is performed over the volume of the star, $\rho
$ is the stellar density.

Equations of motion for the coefficients $b(t)$ can be found e.g.
in Ivanov $\&$ Papaloizou (2004)  as
\begin{equation}
\ddot b_n+\omega^2_0b_n+2i\omega^1_n\dot b_n=S,  \quad  {\rm with }  \quad S=\int d^3x\rho
({\mbox{\boldmath $\xi$}}^{*}_n \cdot \nabla U), \label{e13}
\end{equation}
where  for a normal mode  with time dependence
through a factor $\exp(-{\rm i}\omega t),$ we set $\omega =\omega_0+\omega^1_n,$ with  $\omega_0$  being
the eigenfrequency for the  non rotating star,  and  the frequency correction due to rotation of the star is
$\omega^1_n$,  which can be represented as (see e.g.
Christensen-Dalsgaard 1998)
\begin{equation}
\omega^1_n=n\beta_r\Omega , \label{e14}
\end{equation}
the values of the coefficient $\beta_r\sim 1$ were calculated by
Saio (1981)  for a number of $l=2$ normal modes of a $n=3$ polytropic
star.

Substituting equation (\ref{e3}) in the expression for the source
term $S$ in (\ref{e13}) we get
\begin{equation}
S=A_nQ, \quad Q=\int_0^{R_*} r^3dr \rho (2\xi_r+6\xi_s). \label{e15}
\end{equation}
The overlap integrals $Q$ have been calculated by PT and Lee $\&$
Ostriker 1986 for a number of polytropic stellar models.

From equation (\ref{e13}) it follows that the Fourier transform of
the amplitude $b_n$, $\tilde b_n$, can be expressed through the
Fourier transform of the amplitude $A_n$ defined in equation
(\ref{e3}), $\tilde A_n$, as
\begin{equation}
\tilde b_n={Q\tilde A_n\over D(\sigma)}, \quad D(\sigma
)=\omega^2_0-(\sigma +i\nu -\omega^1_2)^2, \label{e16}
\end{equation}
where we add a small imaginary correction to $\sigma$, $\sigma
\rightarrow \sigma +i\nu$, $\nu > 0$ according to the Landau
prescription and neglected $(\omega^1_2)^2$ as a small quantity.

\section{The energy transfer}\label{energytr}

During the planet's flyby of the star the energy is deposited in
the star at a rate determined by work done by the tidal force
\begin{equation}
\dot E=\int d^3x\rho (\dot {\mbox{\boldmath $\xi$}}\cdot \nabla U), \label{e17}
\end{equation}
and, accordingly, the total energy transfer, $\Delta E$, is
obtained by integration of (\ref{e17}) over time
\begin{equation}
\Delta E=\int^{+\infty}_{-\infty} dt \int d^3x\rho (\dot {\mbox{\boldmath $\xi$}}
\cdot \nabla U), \label{e18}
\end{equation}
where the limits of integration are formally extended to
$\pm\infty$ since the integral over time in (\ref{e18}) is highly
peaked near $t=0$ corresponding to the periastron passage.

We substitute equations (\ref{e3}), (\ref{e11}) and (\ref{e12}) in
(\ref{e18}) and integrate over the volume of the star using the
known orthogonality properties of the spherical functions  and
relations (\ref{e5}) to get
\begin{equation}
\Delta E=Q\sum_{n}\int^{+\infty}_{-\infty} dt \dot b_n A_n^*,
\label{e19}
\end{equation}
where we recall  that the overlap integral $Q$ is given by
(\ref{e15}). Now we express $\dot b_n$ and $A_n$ in (\ref{e18}) in
terms of their respective Fourier transforms using (\ref{e6}) and
integrate the result over time with help of the well known
property that $\int dt e^{-it(\sigma -\sigma^{'})}=2\pi \delta
(\sigma -\sigma^{'})$ to obtain
\begin{equation}
\Delta E=-2\pi i Q^2\sum_n \int ^{\infty}_{-\infty}d\sigma \sigma {|\tilde
A_n|^2\over D(\sigma)}, \label{e20}
\end{equation}
where $D(\sigma)$ is defined in equation (\ref{e16}).

A simple analysis shows that in the limit of $\nu \rightarrow 0$
the integral in (\ref{e20}) is mainly determined by contributions
close to the poles of the expression under the integral, where
$D(\sigma ) \approx 0$ and, accordingly,
\begin{equation}
\sigma \approx \omega^{\pm}_n\equiv \pm \omega_0 +\omega^1_n
\label{e21}
\end{equation}
Taking into account only these contributions in (\ref{e20}) we get
\begin{equation}
\Delta E=(\pi Q)^2\sum_n \left(1+ {\omega_n^1\over \omega}\right) |\tilde
A_n(\omega^{+}_n)|^2+\left(1- {\omega_n^1\over \omega}\right) |\tilde
A_n(\omega^{-}_n)|^2. \label{e22}
\end{equation}
It is convenient to use only positive values of the resonant
frequencies in (\ref{e22}). Taking into account that
$\omega^{1}_{-n}=-\omega^1_{n}$ and, accordingly
$\omega_n^{-}=-\omega_{-n}^{+}$, and using equation (\ref{e7}) we
have
\begin{equation}
\Delta E=(\pi Q)^2\sum_n \left(1+{\omega_n^1\over \omega}\right)|\tilde
A_{n}(\omega^{+}_{n})|^2+\left(1-{\omega_n^1\over \omega}\right)|\tilde
A_{-n}(\omega^{+}_{-n})|^2. \label{e23}
\end{equation}

Now we substitute equation (\ref{e10a}) in (\ref{e23}) to obtain
\begin{equation}
\Delta E=4\left({GMQ\over \Omega_{orb} R_{min}^{3}}\right)^2\epsilon ,
\label{e24}
\end{equation}
where
$$\epsilon=|a_{0}(\omega )|^2+\left(1+{\omega_1^1\over
\omega}\right)|a_1(\omega^+_1)|^2+\left(1-{\omega_1^1\over
\omega}\right)|a_{-1}(\omega^+_{-1})|^2$$
\begin{equation}+\left(1+{\omega_2^1\over
\omega}\right)|a_2(\omega^+_2)|^2+\left(1-{\omega_2^1\over
\omega}\right)|a_{-2}(\omega^+_{-2})|^2, \label{e25}
\end{equation}
and we take into account that the terms with opposite signs of $n$
give the same contribution in the summation series in (\ref{e23}).
Using equation (\ref{e10a}) we can also obtain explicit
expressions for the quantities $|a_n(\omega)|^2$ entering in
(\ref{e25}) in the form
$$|a_n(\omega)|^2=W_2^2(d^2_{n,2}I^{2}_{2,-2}
+d^{2}_{n,-2}I^2_{2,2}+2d_{n,2}d_{n,-2}I_{2,-2}I_{2,2} \cos
4\gamma )$$
\begin{equation}
+2W_2W_0d_{n,0}I_{2,0}\cos 2\gamma
(d_{n,2}I_{2,-2}+d_{n,-2}I_{2,2})+W_{0}^2d_{n,0}^2I^{2}_{2,0},
\label{e26}
\end{equation}
where all $I_{2,n}$ are assumed to be functions of $\omega $ and
we set, for simplicity, $d_{n,m}\equiv d^{(2)}_{n,m}$. One can
easily check that when $\beta=0$ and, accordingly,
$d_{n,m}=\delta_{n,m}$ the expression (\ref{e24}) is equivalent to
the expression (59) of Ivanov $\&$ Papaloizou (2004).

It is instructive to express natural units in (\ref{e24}). Taking
into account that according to our normalisation condition
(\ref{e12a}) the overlap integral $Q$ can be expressed as
$Q=R_{*}M_{*}^{1/2}\tilde Q$, where $R_{*}$ is the radius of the
star and $\tilde Q $ is dimensionless. Let us also introduce
natural energy and frequency scales associated with the star,
$E_{*}=GM_{*}^{2}/R_{*}$ and $\Omega_{*}=\sqrt{GM_*/
R_{*}^{3}}$, respectively, and use the Press Teukolsky parameter
$\eta $ characterising a strength of the tidal interaction,
\begin{equation}
\eta=\Omega_*/\Omega_{orb}={1\over \sqrt{1+q}}\left({R_{min}\over
R_{*}}\right)^{3/2}. \label{e27}
\end{equation}
With help of these new variables the expression (\ref{e24}) can be
rewritten in the form
\begin{equation}
\Delta E=4\left({{q\tilde Q}\over (1+q)\eta}\right)^2\epsilon E_{*} .
\label{e28}
\end{equation}

\section{Numerical solutions of  linear  tidal problem for orbital planes
 inclined to the equatorial plane of the star}\label{Numerical}
 Here we describe numerical solutions for tidal encounters of a planet 
 with a star when the orbit is not coplanar with its equatorial plane.
 We obtain the energy and $Z$ component of the angular momentum  
exchanged due to tides acting on the star. This in turn can be related to the rate of orbital circularisation
starting from high eccentricity (Papaloizou \& Ivanov 2007).

 The numerical procedures follow closely those described in Papaloizou \& Ivanov (2010)
that were applied to tidal encounters with a polytrope of index $n=1$
but with appropriate modifications to consider a polytrope with $n=3.$
These are described below. Only coreless models are considered here.
As in Papaloizou \& Ivanov (2010) grid resolutions of $200\times 200$ and
$400\times 400$ have been considered.

\subsection{A simple stellar model}
We adopt a simple model corresponding to a spherically symmetric 
polytrope of index $n=3.$ It is in hydrostatic equilibrium
such that
\begin{equation}
\frac{1}{\rho}\frac{dP}{dr}=\frac{d\psi}{dr},  \label{eq hydro}
\end{equation}
where $\rho$ is the density, $P=K\rho^{1+1/n}$ is the pressure, with $K$
being the polytropic constant  and $\psi$ is the
gravitational potential arising from the stellar matter. This satisfies
the Poisson equation
\begin{equation}
\frac{ d}{dr} \left(r^2\frac{d\psi}{dr}\right)= -4\pi G\rho.  \label{eq hydropot}
\end{equation}

\begin{figure}
\hspace{-0.5cm}
\subfigure[]{\label{fig:1}
  \includegraphics[width=0.55\textwidth,height=8cm]{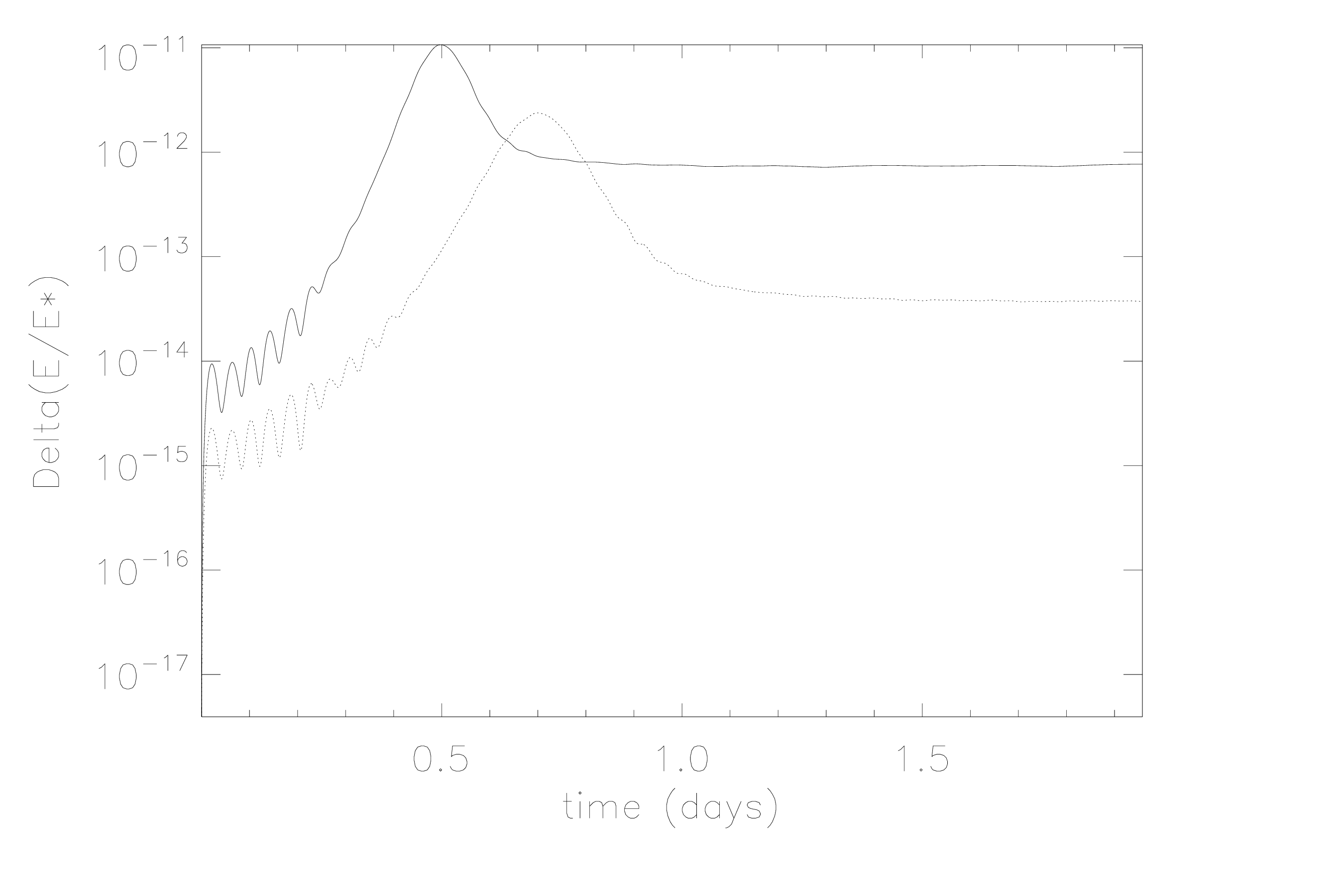}}
\hspace{-1.0cm}
\subfigure[]{\label{fig:2}
  \includegraphics[width=0.65\textwidth,height=8cm]{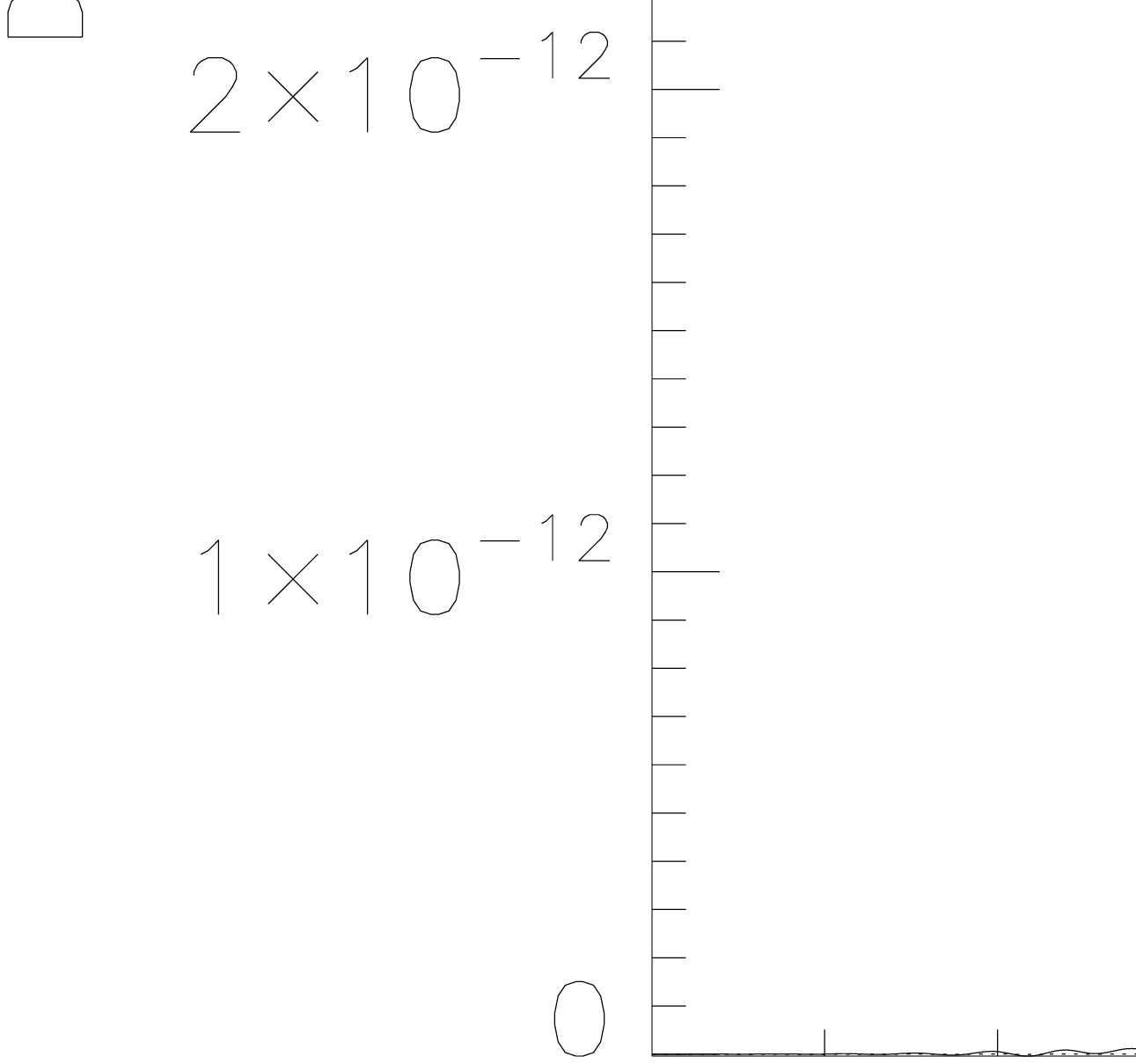}}
\caption{ a) The energy,  in units of $E_*,$ transferred to the non rotating  star as a result of a parabolic
encounter with $\eta = 4\sqrt{2}$ (upper solid curve) and $\eta=8$ (lower dotted curve).
The mass ratio  $q=10^{-3}.$\newline
b) The $z$ component of the  angular momentum,  in units of $J_* = M_*\sqrt{GM_* R_*},$  transferred to the non
rotating  star 
 for $\eta = 4\sqrt{2}$ (upper solid curve) and $\eta=8$ (lower dotted curve).
The mass ratio  $q=10^{-3}.$}
% \label{fig:1}       % Give a unique label
\end{figure}

%\begin{figure}
%\includegraphics[width=\textwidth]{DeltaJnonrot.ps}
%\caption{The angular momentum,  in units of $J_* = M_*\sqrt{GM_* R_*},$  transferred to the non 
%rotating  star as a result of a parabolic
%encounter with $\eta = 4\sqrt{2}$ (solid curve) and $\eta=8$ (dotted curve).
%The mass ratio  $q=10^{-3}.$}
%\end{figure}

\subsection{Linearised equations of motion}
We assume that the star  rotates with uniform angular velocity
$\bmth {\Omega }$ directed parallel to the $Z$ axis.
The hydrodynamic equations for  adiabatic perturbations
induced by an orbiting planet  in a  frame corotating with the star are
\begin{equation}
\frac{\partial {\bf  v}} {\partial t }+2{\bmth {\Omega}}\times{\bf v}
 =-\frac{P^{1/\gamma}}{\rho}\nabla (P'/P^{1/\gamma}) -\xi_r\omega_{BV}^2{\bf e}_r+\nabla U +\nabla\psi'+ \frac{{\bf f}_{\nu}}{\rho}, \label{eq p3}
\end{equation}
where the Brunt Vaisala frequency $\omega_{BV}$ is given by
\begin{equation}
\omega_{BV}^2=  \frac{1}{\rho}\frac{dP}{dr}\left(\frac{1}{\rho} 
\frac{d\rho}{dr} -\frac{1}{\gamma P}\frac{dP}{dr}\right) ,  \label{eq p4}
\end{equation}
${\bf v}= (v_r, v_{\theta}, v_{\phi})$ is the Eulerian velocity perturbation,
$P' $ is the Eulerian
pressure perturbation,   ${\bf f}_{\nu} $ is the viscous
or diffusive force per unit volume, $\psi'$ is the perturbation of the gravitational
potential due to the star  and  as indicated above $U$ is
the external forcing  tidal potential. The adiabatic exponent $\gamma$ is taken to be 
constant and equal to $5/3.$   The velocity perturbation ${\bf v}$
is related to
${\bmth{ \xi}}= (\xi_r, \xi_{\theta},\xi_{\phi}), $  the Lagrangian displacement vector, through
\begin{equation}
\frac{\partial {\bmth{\xi}}} {\partial t } = {\bf v}\label{xidot}.
\end{equation}
 The linearised continuity equation gives
\begin{equation}
\rho^{'}=-\nabla \cdot (\rho {\bmth {\xi}}),  \label{eq p6}
\end{equation}
where $\rho'$ is the density perturbation.
This together with the adiabatic condition gives
 the Eulerian  pressure perturbation as
 \begin{equation}
P'=-\xi_r \frac{dP}{dr}-\gamma P \nabla \cdot (\rho {\bmth {\xi}}).  \label{eq p61}
\end{equation}
 The gravitational potential perturbation $\psi'$ satisfies the linearised Poisson equation
 \begin{equation}
\nabla^2 \psi' = -4\pi G \rho' . \label{eq p62}
\end{equation}
 Equation (\ref{eq p62}) may be solved by expanding the right hand side
 in a series of spherical functions. We simplify the problem by noting that
 we consider solutions  the tidal problem by combining 
 solutions of equation (\ref{eq p3}) where the angular dependence of $U$ 
  is through a single spherical function. To evaluate $\psi'$ we retain only the spherical
  function associated with the forcing in the decomposition of $\rho'.$
  This is an exact procedure in the non rotating case and should be a reasonable
  approximation in the rotating case as  the potential perturbations associated with
  high order spherical functions are expected to be small.
  It amounts to retaining only the quadrupole components of the gravitational
  potential for the problem on hand, a procedure that has been used  successfully in problems
  of stellar dynamics (eg. Allen et al. 1990). 
 
Note also hat the centrifugal term is absent in equation  $(\ref{eq p3})$
being formally incorporated into the potential governing
the static equilibrium of the unperturbed star
and  there is no unperturbed motion in the rotating frame.
As in our previous work (eg. Papaloizou $\&$ Pringle 1981,  Papaloizou \& Ivanov 2010)
we shall neglect centrifugal distortion of the basic equilibrium  which 
enables us to adopt a spherically symmetric unperturbed  model
and density distribution. 
We consider prograde encounters, retrograde encounters and encounters for which the 
orbital plane is at right angles to the stellar equatorial plane.

%\subfigure[]{\label{fig:1}
%  \includegraphics[width=0.6\textwidth,height=8cm]{DeltaEnonrot.ps}}
%\subfigure[]{\label{fig:2}
% \includegraphics[width=0.6\textwidth,height=7.975cm]{DeltaJnonrot.ps}}

\begin{figure}
\hspace{-0.5cm}
\subfigure[]{\label{fig:7}
  \includegraphics[width=0.55\textwidth,height=8cm]{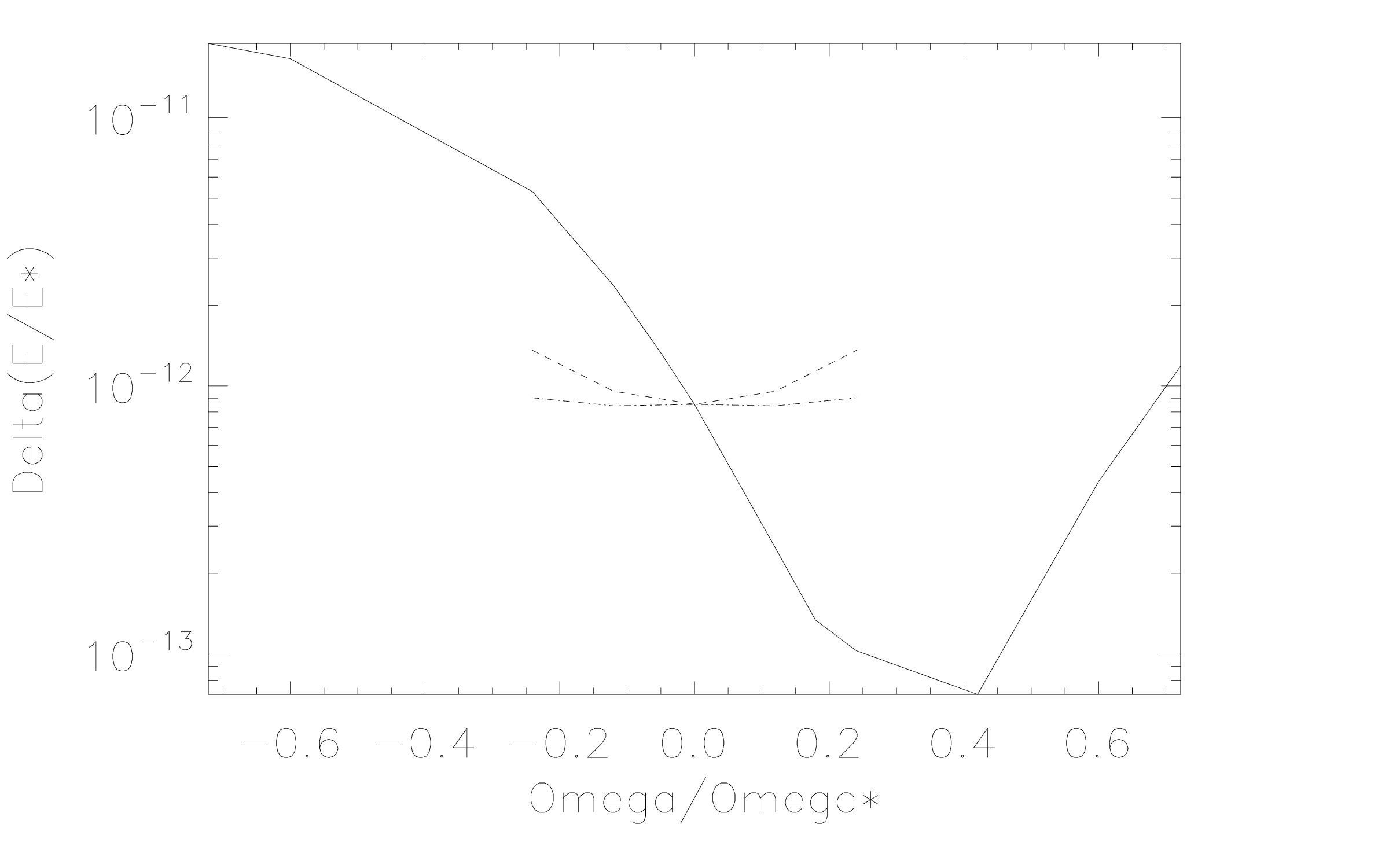}}
\hspace{-1.0cm}
\subfigure[]{\label{fig:8}
  \includegraphics[width=0.65\textwidth,height=8cm]{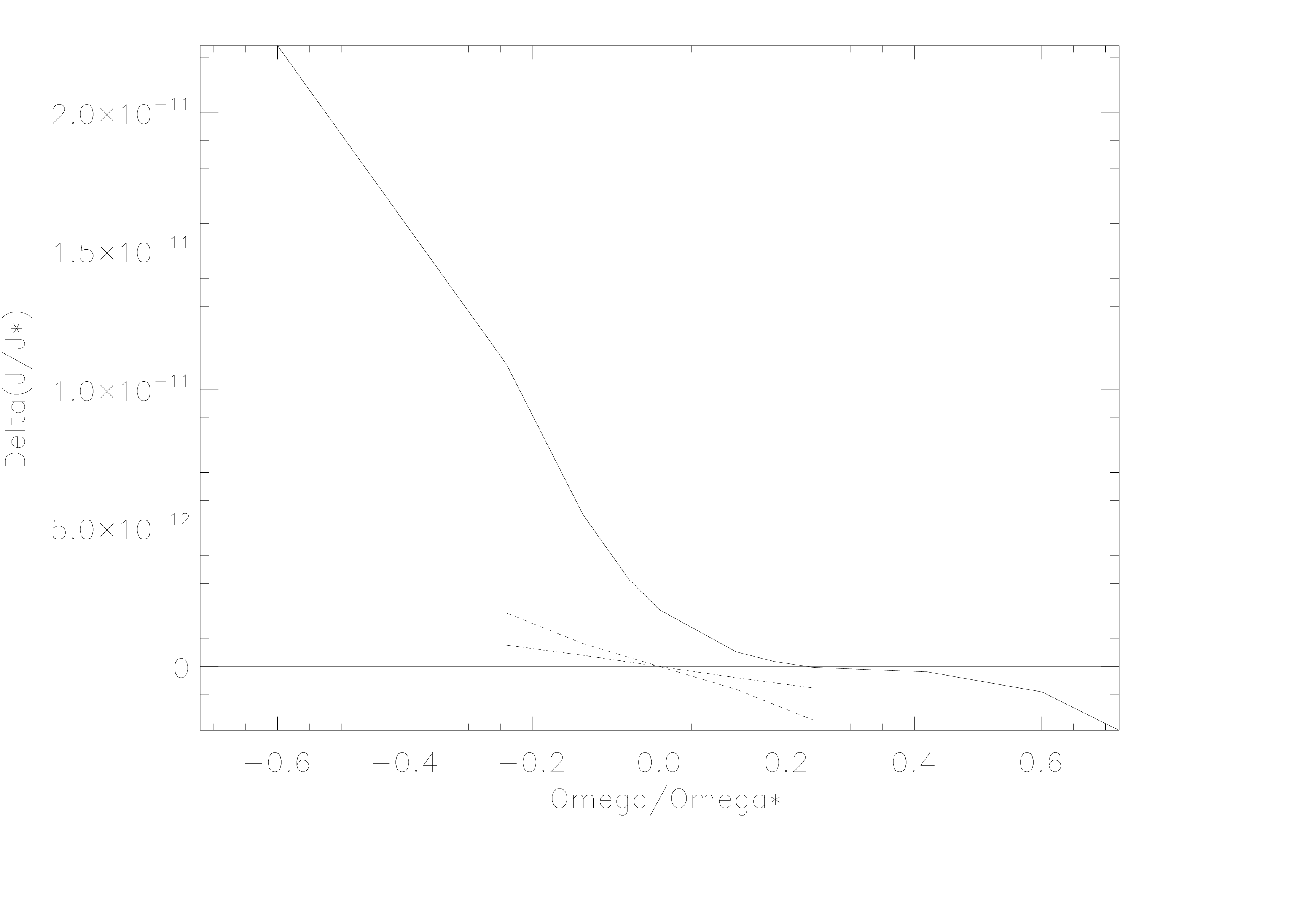}}
\vspace{0.5cm}
\hspace{-0.5cm}
\subfigure[]{\label{fig:9}
  \includegraphics[width=0.55\textwidth,height=8cm]{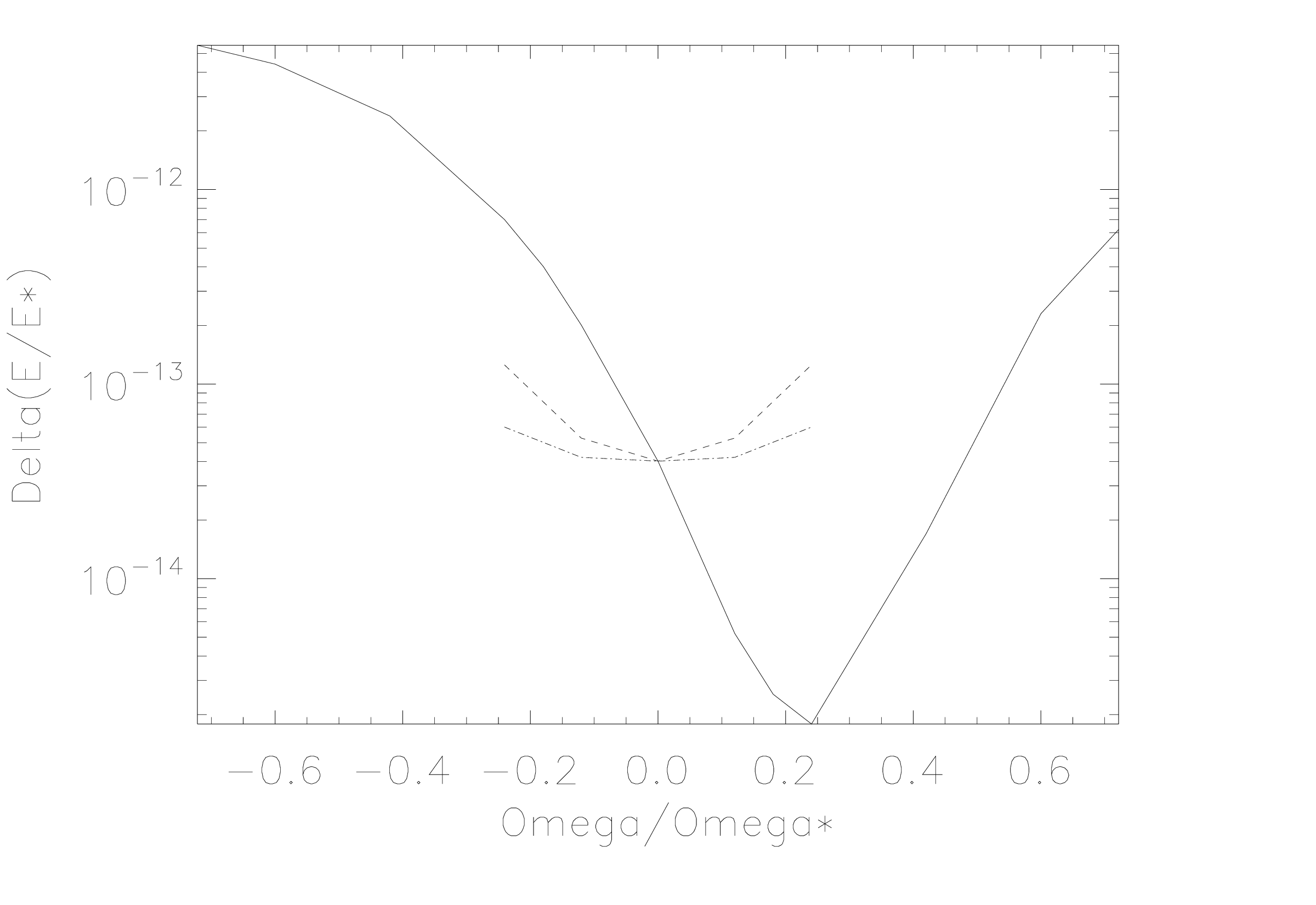}}
\hspace{-1.0cm}
\subfigure[]{\label{fig:10}
\includegraphics[width=0.65\textwidth,height=8cm]{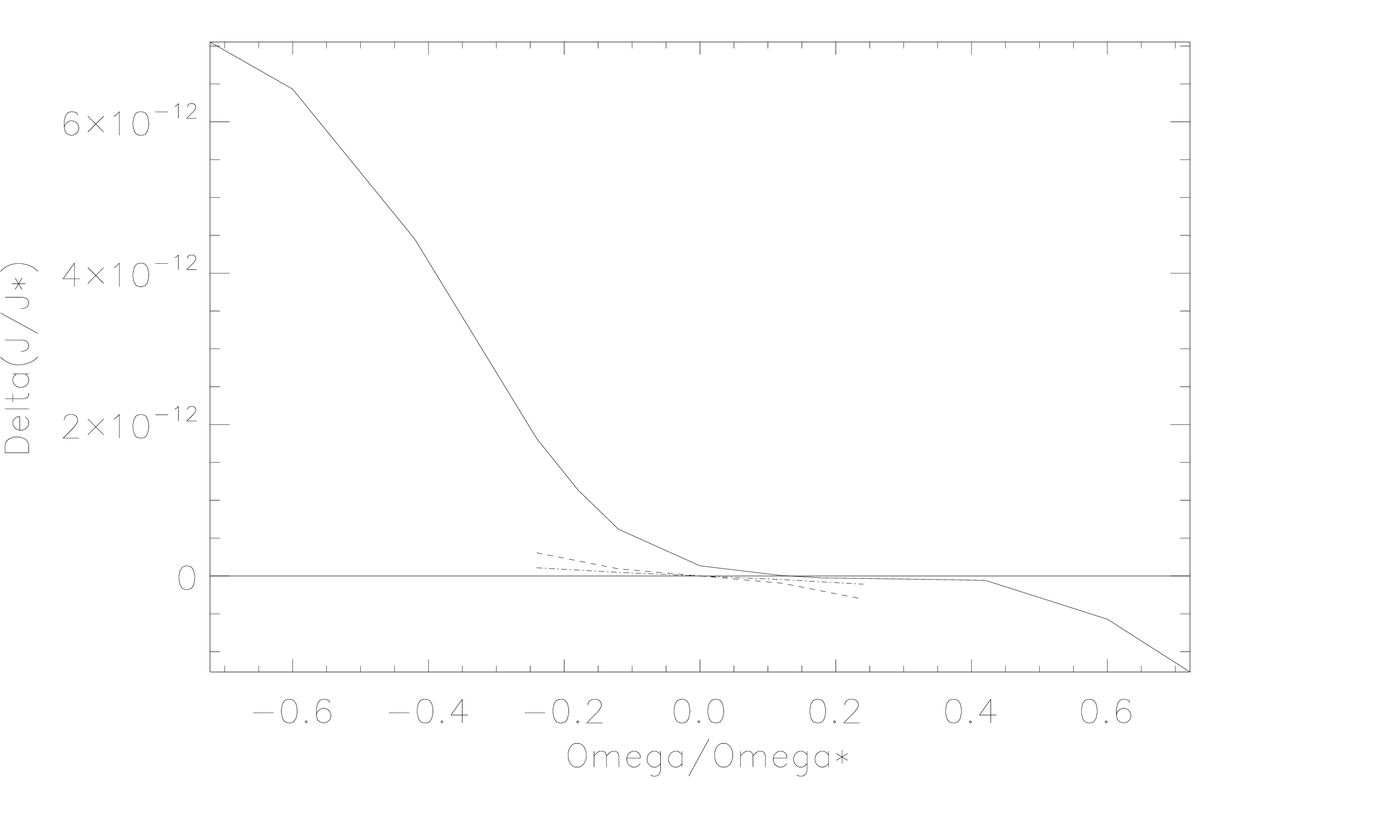}}

\caption{ a) The energy,  in units of $E_*,$ transferred to the star 
for  $\eta = 4\sqrt{2}$ and $q=10^{-3}$  plotted as a function of $\Omega/\Omega_*.$
Negative values of the latter correspond to retrograde rotation.
The upper dashed curve spanning $(-0.25,0.25)$
shows the energy transferred  when the  orbital plane is at right angles to the
stellar equatorial plane where pericentre is  located.
The lower dot dashed curve spanning $(-0.25,0.25)$
is for the same parameters except that  pericentre is  located on the stellar rotation axis.
b) The angular momentum,  in units of $J_* = M_*\sqrt{GM_* R_*},$ transferred to the star. 
The  initially uppermost 
 dashed curve spanning $(-0.25,0.25)$  shows the  $z$ component of the angular momentum transferred 
  when the orbital plane  is at right angles to the
stellar equatorial plane where  pericentre being is  located. The 
initially lowermost  dot dashed curve spanning $(-0.25,0.25)$
corresponds to the same parameters except that  pericentre is  located on the stellar rotation axis.
c) As in a) but for $\eta=8.$
d) As in b) but for $\eta=8.$
}
%\label{fig:7}       % Give a unique label
\end{figure}

\subsection{Tidal forcing potential for prograde and retrograde encounters} 
For prograde encounters the general expression
(\ref{e3}) for the tidal potential reduces to the specific form
\begin{equation} U = {\cal R}\left[\frac{3G M}{4R}\left(\frac{r}{R}\right)^2
  \sin^2(\theta)\exp( 2i(\phi- \Phi)-\frac{ GM}{4R}\frac{r^2}{R^{2}}(3\cos^2\theta-1)\right], \label{Tidpot}
\end{equation}
where ${\cal{R}}$ indicates that the real part is to be
taken and $R(t)$ and $\Phi(t)$ are given by  (\ref{paroR}) - (\ref{parot}). 
The same expression may be used for a retrograde encounter
 by making the replacement $\Phi \rightarrow -\Phi.$
\subsection{Tidal forcing potential for encounters with $\beta = \pi/2$}
In this case when pericentre is in the equatorial plane 
 we may write
\begin{eqnarray}
U&=&\frac{GM}{4R}\left(\frac{r}{R}\right)^2{\cal{R}}
\left[-3\sin^2\theta\cos^2\Phi\exp(2i\phi)\right.\\ \nonumber
 & &\left.-6{\rm i}\sin\theta\cos\theta\sin2\Phi\exp(i\phi)
+(3\cos^2\theta -1)(3\sin^2\Phi-1)\right]
\end{eqnarray}
and when it is on the rotation axis we may write
%\newpage
\begin{eqnarray}
U&=&\frac{GM}{4R}\left(\frac{r}{R}\right)^2{\cal{R}}
\left[-3\sin^2\theta\sin^2\Phi\exp(2i\phi)\right.\\ \nonumber
 & &\left.+6{\rm i}\sin\theta\cos\theta\sin2\Phi\exp(i\phi)
+(3\cos^2\theta -1)(3\cos^2\Phi-1)\right].
\end{eqnarray}
We remark that changing the direction of motion
along the orbit by the replacement $\Phi \rightarrow -\Phi$
does not change the energy transfer or the change in the $Z$ component of angular 
momentum. This corresponds to encounters with $\beta=\pi/2$ and $\beta=3\pi/2$
giving the same results for these quantities.
The above tidal potentials consist of  a sum of terms with different azimuthal mode number $m.$
These can be considered independently and the energy and angular momentum transferred
as a result of the action of each term
evaluated as the  canonical energy and angular momentum
 obtained after the encounter is over (see below).
The results can then be  added  to get the total. 
As in Papaloizou \& Ivanov (2010) the encounters are started 
with the perturber at eight times the pericentre distance
with $R$ and $\Phi$ determined from (\ref{paroR}) - (\ref{parot}).  It can be seen that
the results are unaffected 
by applying an arbitrary complex phase shift to each individual forcing term.

\subsection{Addition of diffusive effects}\label{sec2.2}

In order to avoid potential numerical problems arising from
the excitation the  a dense or continuous spectrum of normal modes 
that may be associated with a rotating star,
we follow Papaloizou \& Ivanov (2010) and incorporate
 a diffusive force per unit mass, ${\bf f_{\nu}}$,  in the simulations.
For models with non zero rotation this is taken to be of the form
\begin{equation}
{\bf f}_{\nu} = \frac{\partial}{\partial r}\left(\rho\nu\frac{\partial {\bf v}}{\partial r}\right)
+\frac{1}{r^{2}}
\frac{\partial }{\partial \mu}\left(\rho \nu\sin\theta \frac{\partial {\bf v}}{\partial \mu}\right),
\end{equation}
where the diffusion coefficient or effective kinematic viscosity
$\nu = \nu_0 \sqrt{GM_*R_*}$ was taken to be a constant. For  simulations
 with  a grid resolution of $200\times 200$
we adopted
$\nu_0=1.3\times10^{-6}$ and for a grid
 resolution of $400\times400$
we adopted $\nu_0=6.5\times 10^{-7}.$
For non rotating models $\nu_0=0.$
We remark that energy and angular momentum transfers measured
by evaluating the canonical energy and angular momentum
just after the encounter has completed
are robust to changes in numerical resolution and diffusivity.

\subsubsection{The canonical energy}

When the Lagrangian pressure perturbation vanishes
at the outer boundary, the canonical energy (defined in the rotating frame) 
appropriate
to forcing through a  Fourier mode with azimuthal mode number, $m,$
 may be written as
\begin{eqnarray}
 E_c&=&0.25(1+\delta_{m,0})\left[ \int_V  \rho\left(   |{\bf v}|^2+\omega_{BV}^2|\xi_r|^2  
 +|P'|^2/(\gamma P)  \right)d\tau\right.\nonumber\\
 &-&\left.\int\frac{ |\nabla\psi'|^2}{4\pi G}d\tau-\int_A|\xi_r|^2(dP/dr) dS\right], \label{canE}
\end{eqnarray}
where $\delta_{m,0})$ is the Kronecker $\delta.$
The canonical energy contains contributions from the pressure
contribution, buoyancy and self-gravity. The last two
were not included in Papaloizou \& Ivanov (2010)
where a polytrope with $n=1$  was considered using  the Cowling approximation.
The  first volume integral is taken over the 
volume of the star $V,$ while the second giving the gravitational energy
is taken over all space. The  surface integral is taken over
the surface area $A.$ The contribution of the surface
term formally vanishes when the density at the surface
is zero. When the density at the surface is relatively small
as in our  model,  this term  gives a negligible contribution.
When tidal forcing operates, the time rate of $E_c$
 gives the rate of energy  transfer to  the star. After the encounter
tidal forcing tails away  and $E_c$ is conserved in the absence of dissipation.
In the presence of dissipation it decays with time.

Let us stress the difference between the energy defined in the rotating frame and in 
the inertial frame. The energy transfer in the rotating frame $\Delta E_{r}=E_{c}$ is 
related to the energy transfer in the inertial frame (see equation (\ref{e28}) and 
the next Section ) as
\begin{equation}
\Delta E =\Delta E_{r}+\Omega \Delta J_{z}, \label{eqnnn}
\end{equation}   
where $J_{z}$ is the transfer of $z$th component of angular momentum, which may be equated 
to the canonical angular momentum, $J_{c}$, contained in the perturbed modes, see below. 
While it is more convenient to operate with $\Delta E_{r}$ in our numerical work, the analytical 
work as well as different astrophysical applications are concerned directly with  $\Delta E$. 
Therefore, we illustrate the behaviour of  $\Delta E_r$ in  this Section when discussing our numerical results
and  $\Delta E $ in the next Section where  numerical and   analytic results are compared.

\subsubsection {The canonical Angular momentum}
The corresponding  canonical angular momentum is given by
\begin{equation}
J_c=-{\cal I} m\left[0.5m\int_V  \rho\left(   \xi_r^*v_r    +\xi_{\theta}^*v_{\theta} +\xi_{\phi}^*v_{\phi}
+ 2\Omega (\xi_r\xi^*_{\phi}\sin\theta+  \xi_{\theta}\xi_{\phi}^* \cos\theta )\right)d\tau\right],
 \end{equation}
where ${\cal I}m$ indicates that the imaginary part is to be taken.
Here we recall that the state variables are complex.
This behaves in a manner  analogous to the canonical energy, but as applied
to the total angular momentum content of the star.

\section {Numerical results}\label{Numres}

\subsection{The energy and angular momentum transfer for a close encounter}

\begin{figure}
\hspace{-0.5cm}
\subfigure[]{\label{fig:3}
  \includegraphics[width=0.55\textwidth,height=8cm]{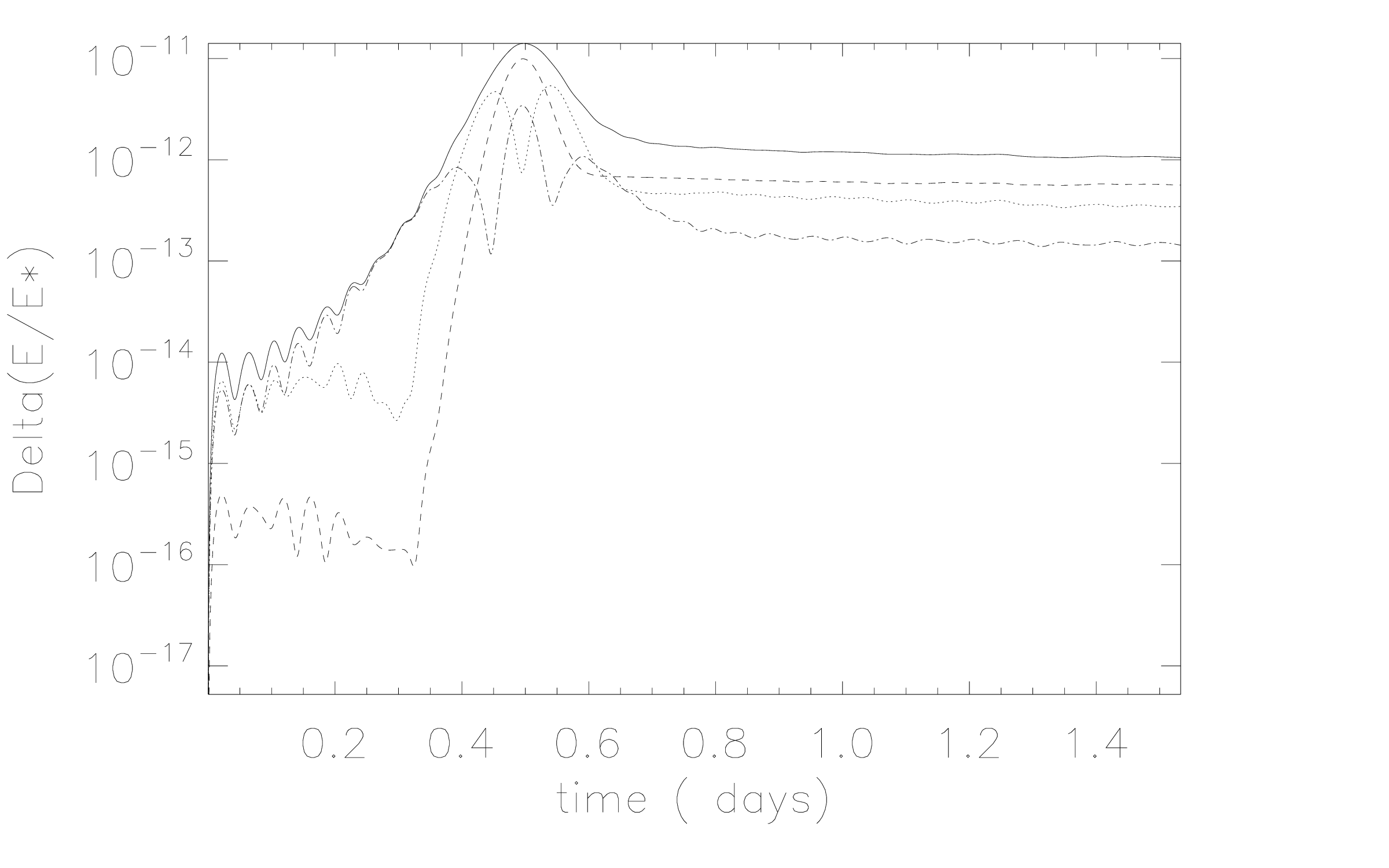}}
\hspace{-1.0cm}
\subfigure[]{\label{fig:4}
 \includegraphics[width=0.65\textwidth,height=8cm]{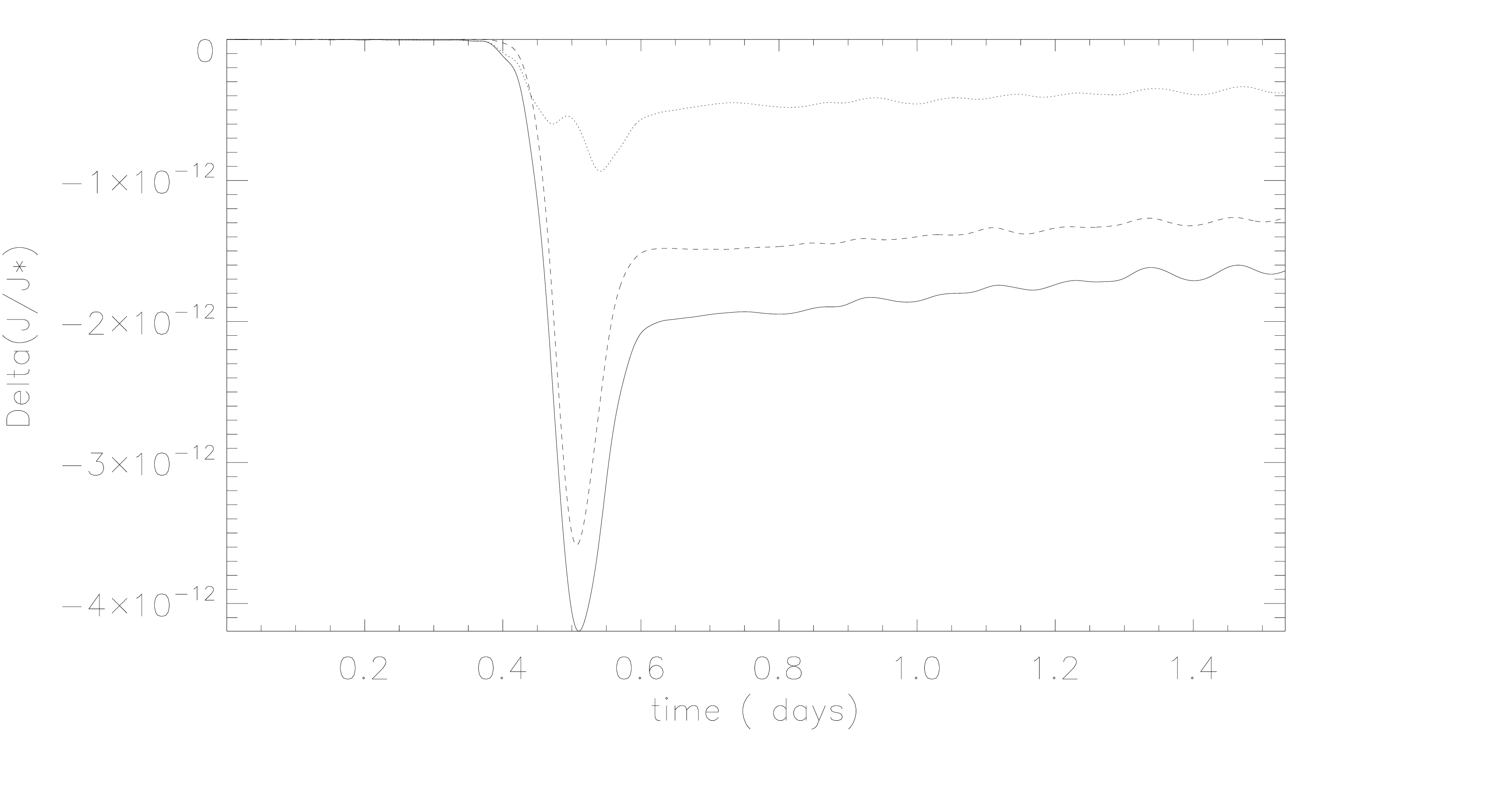}}
\vspace{0.5cm}
\hspace{-0.5cm}
\subfigure[]{\label{fig:5}
 \includegraphics[width=0.55\textwidth,height=8cm]{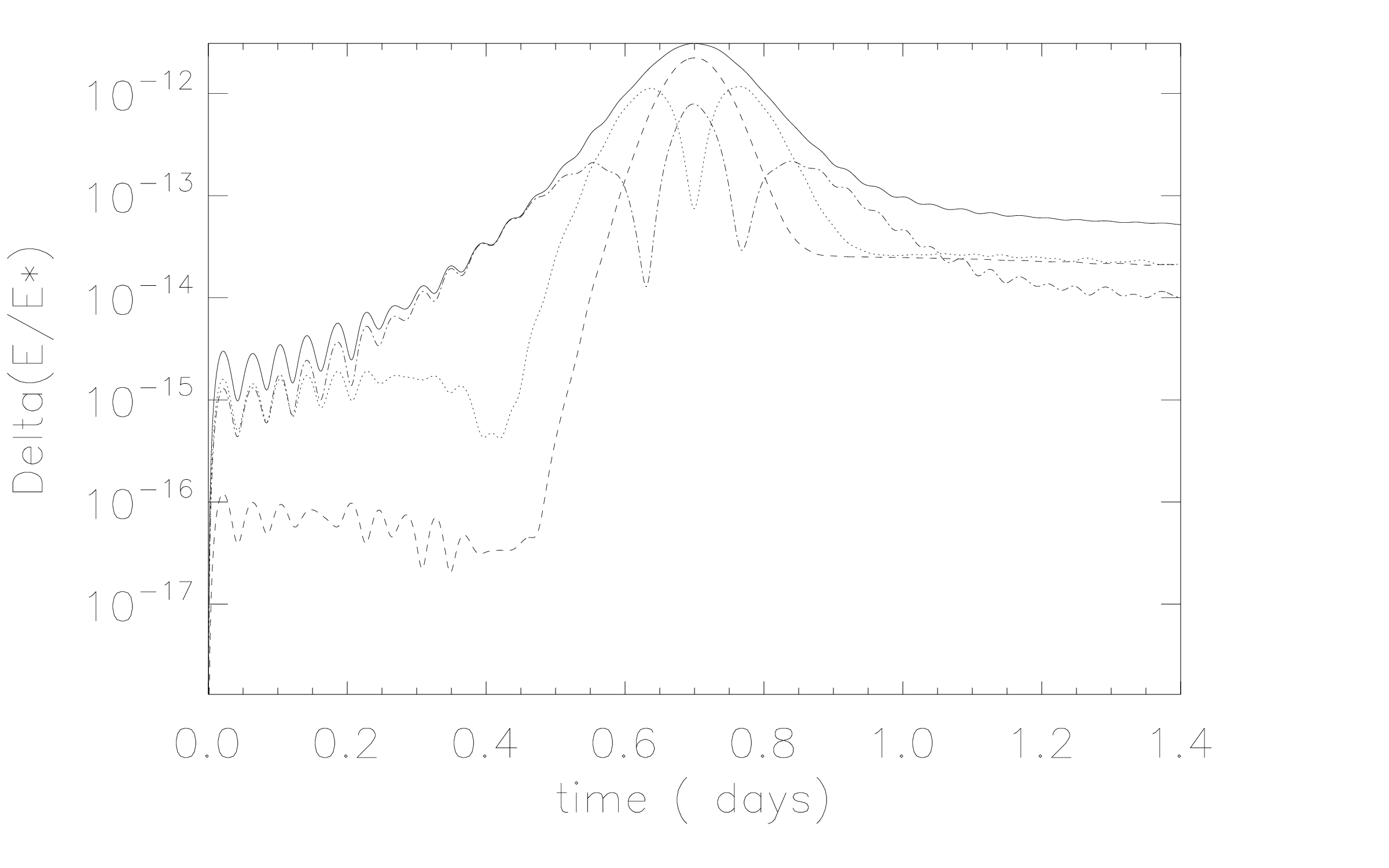}}
\hspace{-1.0cm}
\subfigure[]{\label{fig:6}
\includegraphics[width=0.65\textwidth,height=8cm]{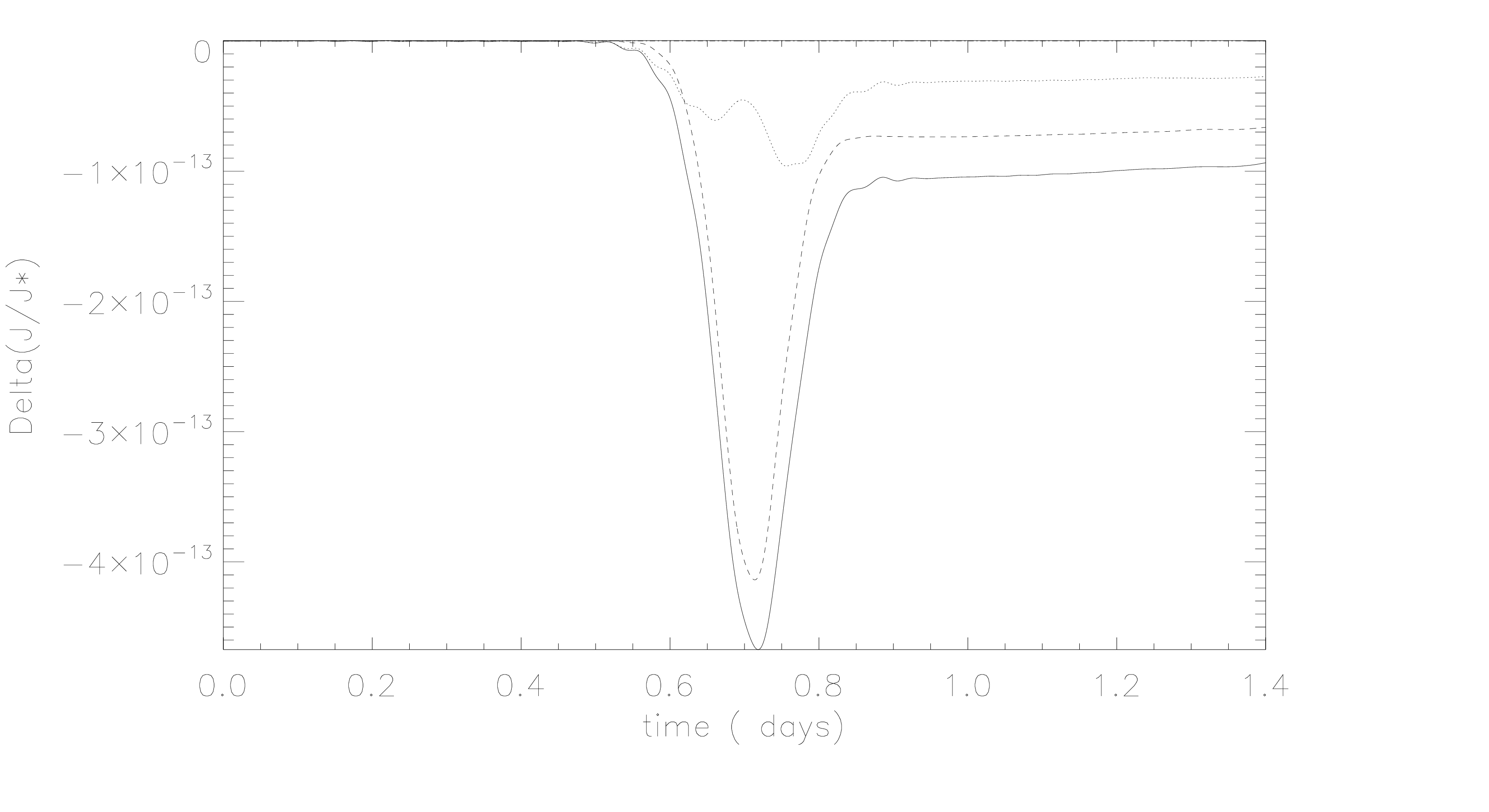}}
\caption{a) The energy,  in units of $E_*,$ transferred to the star for
 $\eta = 4\sqrt{2},   q=10^{-3}$   and  $\Omega/\Omega_*= 0.24$
(uppermost solid curve). The orbital plane was at right angles to the stellar equatorial plane
with pericentre located in the latter plane.
The initially lowermost dashed curve shows the contribution from terms in the forcing potential
with $m=2.$ Similarly the  initially next lowermost dotted curve corresponds to $m=1$
and the ultimately  lowermost dot-dashed curve to $m=0.$
b) The $z$ component of the angular momentum ,  in units of $J_*,$ transferred to the star 
(lowermost solid curve). 
The ultimately intermediate dashed curve shows the contribution from terms in the forcing potential
with $m=2.$ The  ultimately uppermost dotted curve corresponds to $m=1.$
c) As in b)  but for $\eta=8$ and $\Omega/\Omega_*= 0.12.$
d) As in c)  but for $\eta=8$ and $\Omega/\Omega_*= 0.12.$ }
%\label{fig:3}       % Give a unique label
\end{figure}

%\begin{figure}
%  \includegraphics[width=0.6\textwidth,height=7cm]{jinclinedeqom1.5eta4sqrt2.ps}
%\caption{The angular momentum ,  in units of $E_*,$ transferred to the star as a result of a parabolic
%encounter with $\eta = 4\sqrt{2}$ and $q=10^{-3}$   for  $\Omega/\Omega_*= 0.24$
%(solid curve). The orbital plane was at right angles to the stellar equatorial plane
%with pericentre located in the latter plane.
%The dashed curve shows the contribution from terms in the forcing potential
%with $m=2.$ Similarly the dotted curve corresponds to $m=1.$}
%\label{fig:4}       % Give a unique label
%\end{figure}

%\begin{figure}
% \includegraphics[width=0.6\textwidth,height=7cm]{INCLINEDOM0.75eta8.ps}
%\caption{As in Fig. \ref{fig:3} but for $\eta=8$ and $\Omega/\Omega_*= 0.12.$ }
%\label{fig:5}       % Give a unique label
%\end{figure}

%\begin{figure}
%  \includegraphics[width=0.6\textwidth,height=8cm]{JINCLINEDOM0.75eta8.ps}
%\caption{As in Fig. \ref{fig:3} but for $\eta=8$ and $\Omega/\Omega_*= 0.12.$ }
%\label{fig:6}       % Give a unique label
%\end{figure}

The energy, in units of $E_*,$  transferred to  a non rotating  star during   parabolic
encounters  with a planet with mass ratio  $q=10^{-3}$ for  $\eta = 4\sqrt{2}$  and $\eta=8$  is shown 
as a function of time in Fig. \ref{fig:1}.
These results are for the  $m=2$  tide which is the dominant  one (Papaloizou \& Ivanov 2004, 2010). 
The corresponding angular momentum transferred to the star is plotted as  a function of time
in Fig. \ref{fig:2}. The energy and angular momentum transferred level of at constant values
after the tidal interaction is essentially complete,
the characteristic time associated with the encounter being about $1$ day.
This corresponds to the excitation of the normal modes associated with the rotating star
which in this case correspond to $f,$  $p$ and  $g$ modes. It is found that $g$ modes up to order
$9$ may be significant (see Lee \& Ostriker 1986). In practice we measure transferred
quantities at $t=0.75 d$ when  $\eta = 4\sqrt{2}$ and at $t=1.4 d$ when $\eta = 8.$
Our results for the final energies transferred to the star
agree with theirs to about the $10\%$ level.

\subsection{Prograde and retrograde encounters with a rotating star}\label{protetnum}

The energy,  in units of $E_*,$ transferred to the star as on completion  of a parabolic
encounter with $\eta = 4\sqrt{2}$ and $q=10^{-3}$  is plotted as a function of $\Omega/\Omega_*$ in Fig. \ref{fig:7}.
Negative values of $\Omega/\Omega_*$  correspond to retrograde rotation.
The angular momentum,  in units of $J_* = M_*\sqrt{GM_* R_*},$ that is  transferred
is plotted as a function of $\Omega/\Omega_*$ in Fig. \ref{fig:8}.
In  Fig. \ref{fig:9} and \ref{fig:10} the corresponding plots are given for $\eta = 8.$ 
The energy transferred decreases as $\Omega$ is increased from zero until a minimum value is attained, after
which it increases  with $\Omega$ again. On the other hand the energy transferred increases with increasing retrograde
rotation attaining values up to more than three orders of magnitude larger than the minimum one
for $\Omega/\Omega_* \sim -0.6.$ when $\eta=8.$ Thus circularisation  from high eccentricity
as a result of stellar tides can be expected to be much more effective for significant retrograde rotation.
For $\eta = 4\sqrt{2}$ the minimum energy transfer occurs for $\Omega/\Omega_*\sim 0.4$ while  for $\eta = 8$
this occurs for $\Omega/\Omega_*\sim 0.25.$ The  actual angular velocity at pericentre $\Omega_p/\Omega_*
= \sqrt{2}/\eta,$ so that  $\Omega_p/\Omega_* =0.25$ and $0.18$ for these two cases respectively.
Accordingly this minimum  transfer occurs when $\Omega\sim 1.5\Omega_p,$ which is approximately the angular velocity
for which the angular momentum transfer reverses sign, this transition occurring
when  $\Omega\sim \Omega_p,$ (see Figs.  \ref{fig:8} and  \ref{fig:10}). 
This characteristic behaviour was also noted  in the case of a polytrope with index $n=1$
and a barotropic equation of state ( Ivanov \& Papaloizou 2004, 2007).

We have also considered encounters for which the orbital plane and the stellar equatorial
plane are perpendicular. Cases for which pericentre was in the equatorial plane
and on the rotation axis have been considered. 
The  energy transferred to the star for some of these cases is plotted in Figs. \ref{fig:7} and  \ref{fig:9}.
The component of angular momentum in the $z$ direction transferred to the star is plotted  in Figs. \ref{fig:8} and  \ref{fig:10}.
The energy transferred in these cases is also seen to significantly exceed the minimum that is transferred in the prograde case.

To look at such encounters more closely, the energy,  in units of $E_*,$ transferred to the star as a result of a parabolic
encounter with $\eta = 4\sqrt{2}$ and $q=10^{-3}$   for  $\Omega/\Omega_*= 0.24$
when the orbital plane was at right angles to the stellar equatorial plane
where pericentre was  located is  plotted as a function of time in Fig. \ref{fig:3}.
The $z$ component of angular momentum transferred is plotted in Fig. \ref{fig:4}.
The   contributions from the  terms in the forcing potential
with $m=2, m=1$
and  $m=0$ are indicated. Clearly the last of these does not affect 
the $z$ component of angular momentum.  Note that significant contributions come from
$m=1$ as well as $m=2.$
The corresponding results for $\eta=8$ and $\Omega/\Omega_*= 0.12$
are shown in  Figs.  \ref{fig:5} and  \ref{fig:6}.

\section{Comparison of numerical and analytic results and circularisation time scales}\label{Numanalcomp}

\subsection{Numerical  versus  analytic results}
\begin{figure}
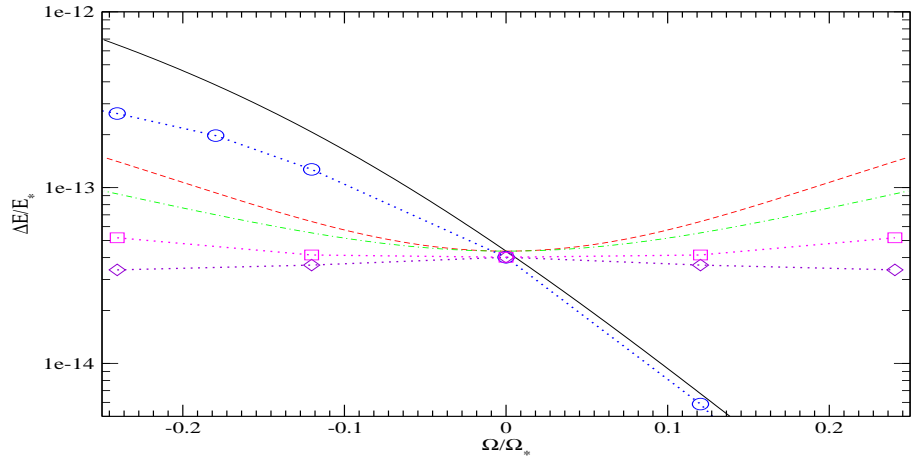

 \vspace{1cm}
\subfigure[\empty]{\label{figg1}
\includegraphics[width=\textwidth,height=6cm]{eom4sq2.eps}}
%\vspace{1.5cm}

\begin{minipage}[b]{1.1\linewidth}\centering
\vspace{1cm}
 \hspace{16cm} { a)}
\vspace{1cm}
 \end{minipage}

%vspace{1cm}
\subfigure[\empty]{\label{figg2}
\includegraphics[width=\textwidth,height=6cm]{eom8.eps}}
%\vspace{3.5cm}
\begin{minipage}[b]{1.1\linewidth}\centering
\vspace{1cm} 
\hspace{18cm}{b)}
\vspace{1cm}
 \end{minipage}
%\vspace{0.5cm}
\caption{ a) Analytic and numerical dependencies of the energy transferred to the star in units of 
the characteristic energy $E_{*}$ ($E_{*}=3.8\cdot 10^{48}ergs$
for a star of  solar  mass and radius) on the angular 
velocity  of the star in units of the characteristic frequency 
$\Omega_*$ ($\Omega_*=6.23\cdot 10^{-4}s^{-1}$ 
for a star of  solar  mass and radius), for $\eta=4\sqrt 2$. 
See the text for a description of the different curves and symbols.
b) Same as a)  but for $\eta=8$.}
%\label{figg1}       % Give a unique label
\end{figure}

%\begin{figure}
%\vspace{1.5cm}
%\includegraphics[width=\textwidth]{eom8.eps}
%\vspace{1cm}
%\caption{Same as Fig (\ref{figg1}) but for $\eta=8$. }
%\label{figg2}       % Give a unique label
%\end{figure}

In order to calculate the analytical value of the energy transfer,
$\Delta E$, we should know the values of the mode eigen frequencies,
the overlap integrals and the frequency splitting coefficients
$\beta_r$. We take into account the $f$-mode, $5$ low order $p$-modes
and $14$ $g$-modes, which give the most important contribution to the
energy transfer. The eigen frequencies and the overlap integral are
taken from the papers Lee $\&$ Ostriker (1986) and Ray et al (1987). The
coefficients $\beta_r$ for the $f$, $p$ and several low order
$g$-modes are taken from the paper Saio (1981). For the rest of
$g$-modes we use the WKBJ results that $\beta_r\approx 1-{1\over
  L(L+1)}$, where $L=2$ in our case, see e.g. Christensen-Dalsgaard
(1998).

In Figs. \ref{figg1} and \ref{figg2} we show the dependencies of the energy transferred to the star as a result of 
the planet flyby on the angular velocity of the star, for $\eta=4\sqrt 2$ and $\eta=8$, respectively\footnote{Let 
us recall  that here we show the energy defined in the inertial frame, which is related to 
the energy defined in the rotating frame discussed in Section \ref{Numres} 
 through equation (\ref{eqnnn}).}. The solid, 
dashed and dot dashed curves represent the analytic results while the symbols  indicate the corresponding 
numerical results. 
The solid curve and circles correspond to the situation when the equatorial and orbital planes coincide 
($\beta =0$).  The dashed curve and squares correspond to  an orbit
with normal to the orbital plane perpendicular to the 
rotational axis and the direction to pericentre laying in the
equatorial plane ($\beta=\pi /2$ and $\gamma 
=\pi /2$). The dot dashed curve and diamonds are calculated for the same $\beta =\pi /2$ but for the direction 
to pericentre being aligned with the rotational axis, and, accordingly, $\gamma =\pi$. Note the one can either consider 
negative values of $\Omega $ as in Figs. \ref{fig:3}-\ref{fig:6},
 or a change of the inclination angle  such that $\beta \rightarrow \beta + \pi$ keeping 
the  value of $\Omega $ positive. Both approaches are physically equivalent and give the same results. In what follows 
we describe the results corresponding to positive (negative ) values of $\Omega $ as prograde 
(retrograde ) and the results corresponding to $\beta=0$ ($\beta=\pi/2$) as planar (perpendicular).
As seen from Figs.  \ref{figg1} and \ref{figg2},
in the planar case the  analytic and numerical  results corresponding to positive rotational 
frequencies are in  good agreement,  while the  results  corresponding to 
negative frequencies  deviate,  leading to 
systematically larger analytic  values of $\Delta E$ as compared to 
 the numerical determined ones. The deviation is of the 
order or smaller than $50\%$ for $-\Omega /\Omega_{*} < 0.1.$
 For the largest  value of  $-\Omega /\Omega_{*} \approx 0.25$ shown, 
it is about a factor of two for $\eta=4\sqrt 2$ and about a factor of $2.5$ for  $\eta=4\sqrt 2.$
 In the perpendicular cases both analytic and 
numerical results show much weaker dependence on $\Omega .$
 The analytic curves  again correspond to  larger values of $\Delta E,$ 
with results deviating most significantly from the numerical ones at
 large values of $|\Omega |.$ The degree of deviation behaves in a  similar 
manner to the planar case.
 Note that the analytic curves always grow with increasing   $|\Omega |$ while the numerical curve
corresponding to $\gamma =\pi/2 $ ($\gamma = 0 $) is weakly increasing (decreasing ) with  $|\Omega .|$
This weak growth (decrease) is probably within numerical errors and should be treated  with  caution.

The origin of the deviation at large values of  $-\Omega /\Omega_{*} $ 
is unclear at present. It cannot be fully 
accounted for by the increasing importance of  non-linear corrections to the rotational 
shift of eigen frequencies given by equation (\ref{e14})  
with increasing  $|\Omega |.$
 We found  that phenomenological non-linear corrections provided by Lai (1997)
do not significantly affect  our analytic  results for the  range of $\Omega$ considered.
 However, for the  values of $\eta$ considered,
the most significant contributions to the energy transfer are provided by g-modes  of order of $9-10.$ 
The eigenfrequencies of these modes being   $\sim 0.5\Omega_*$ 
are  a factor of $\sim 2.5$ larger than the maximum
angular velocity considered. Thus they are approaching the so called inertial
regime for which eigenfrequencies are less than $|2\Omega|$ in magnitude.
 It is possible  that the overlap integrals  become  significantly modified in such a situation. 
This issue will be  explored  in  future work.

\subsection{The time scale for the initial stage of  tidal circularisation}\label{circtime}

\subsubsection{A basic approach}

In principle, the results reported above can be used in numerical experiments  in which  a system of gravitationally 
interacting planets is considered,  it being  assumed that once a  planet has  obtained a sufficiently small value 
of its orbital angular momentum as a result of planet-planet gravitational scattering, 
it can lose orbital energy due to tidal interaction  with the central star during successive periastron passages. 
If the effect of tidal interaction on changing the
orbital energy 
becomes  much larger  than residual  effects associated with
planet-planet scattering, the planet is said to be tidally captured by the star.
After being tidally captured,  the planet
interacts mainly with the central star due to dynamic and quasi-static tides exerted in itself and in the star.
Under certain conditions discussed in the next section the orbital
energy decreases,  either  systematically or on average,
while  the orbital period and orbital eccentricity  also decrease.
In this way it is possible to account for  at least some of the so-called ``Hot Jupiters''
 - giant  planets with orbital periods ranging from a fraction
of a day to several days.      
 
The central star can rotate, with its rotation axis inclined  with respect to  the normal to the 
orbital plane by an angle  $\beta,$  after the tidal capture.  This angle  is
determined by the  particular sequence of  planet-planet interactions
that led to the tidal capture.
Since  retrograde tidal interactions that transfer
energy to the star  are stronger than  prograde ones,
while the angle $\beta $ may be random or have 
a weak correlation with the direction of stellar rotation,
this property of the tidal interaction  may influence 
the observed relative  number of  Hot Jupiters on retrograde and prograde  orbits.
Thus it is important to estimate the potential magnitude of this tidal effect. 

As Hot Jupiters may have been formed when the star was young,
in making such an estimate it is  very important to account for the fact  that the
angular velocity  of the star, $\Omega,$  is a function of stellar age.
This function can be a rather complicated and is expected 
to be different for stars with different masses and formation 
histories (e.g. Scholz 2009).
In this paper we  make only  preliminary qualitative estimates of 
the importance of tides exerted in a rotating star on the formation of Hot Jupiters.
Accordingly, we use  the simplest
plausible form of this function known as the Skumanich law (Skumanich 1972), 
which states that the angular velocity  is
inversely proportional to square root of the stellar age.
Thus we  adopt the following expression for the stellar angular velocity:
\begin{equation}
\Omega=2.4\cdot 10^{-6}\sqrt{{5\cdot 10^9yr\over t}}s^{-1}. \label{eqn1}
\end{equation}       
Here $\Omega$ is normalised   so that the star has a  rotation  period of one  month at an age of 
$t=5\cdot 10^9yr$.

In this paper we do not 
consider the planet-planet interaction stage, leaving this for a future work. Here we 
assume that tidal capture has happened and the planet interacts only with the star. 
It supposed to be initially on a 
highly eccentric ( essentially parabolic)\footnote{A planet with orbital parameters typical for the 
stage of tidal interaction right after the tidal capture would have its periastron distance of 
order of several (say, $4$) stellar radii, while its semimajor distance would be of order of
a few (say, $10$) astronomical units (e.g. Nagasawa et al 2008). For such parameters orbital eccentricity 
$e$ is of order of  $0.998$ and approximation of tidal encounters as happening on parabolic flybys is 
well justified.} orbit with well defined values of 
 the angles $\beta $ and $\gamma $ and have  
a value of the orbital angular momentum corresponding to a given value of the parameter $\eta $ defined in equation 
(\ref{e27}). The orbital energy, eccentricity and period are gradually changing their values due to tidal interactions on 
some characteristic evolution time scale, $t_{ev}$, specified below.  
As discussed in e.g. Ivanov $\&$ Papaloizou (2004,2007),  orbital evolution as a result of tides
acting in the planet approximately conserve orbital angular momentum
because the planet spins up (or slows down) by tides to a state of
rotation corresponding to the so-called pseudo-synchronisation, where
the angular momentum exchange between the orbit and pulsational modes
of the planet is absent. This happens on a quite short time scale due
to a small moment of inertia of the planet compared to that of the
orbit.

In the case of tides exerted in the star we have an opposite
situation where the stellar moment of inertia is typically  much larger than the
orbital one. Therefore, the star cannot be brought by action of tides
to the state of pseudo-synchronisation during the evolution of the
system, and some amount of angular momentum is transferred from the orbit to the 
star in the course of it thus decreasing the value of the orbital
angular momentum. We would like to estimate this value assuming that tides
are effective all the way down to smallest values of the orbital
eccentricity. Let us stress again that the dynamic tides discussed
in this paper are effective only at sufficiently large values of 
the eccentricity $e\sim 1$ ( see the next section).
 However, it seems plausible to suppose that 
 processes not considered here, such as the nonlinear  breaking of excited waves,
 critical latitude phenomena in the planet,  or
friction, resulting from for example turbulence,  acting on quasi-static tides  could be important
when the eccentricity is moderate and small, and 
that they can eventually bring the system to the state having a low
eccentricity $e\sim 0$ typical for the Hot Jupiters (e.g. Barker $\&$ Ogilvie 2009). 

\begin{figure}
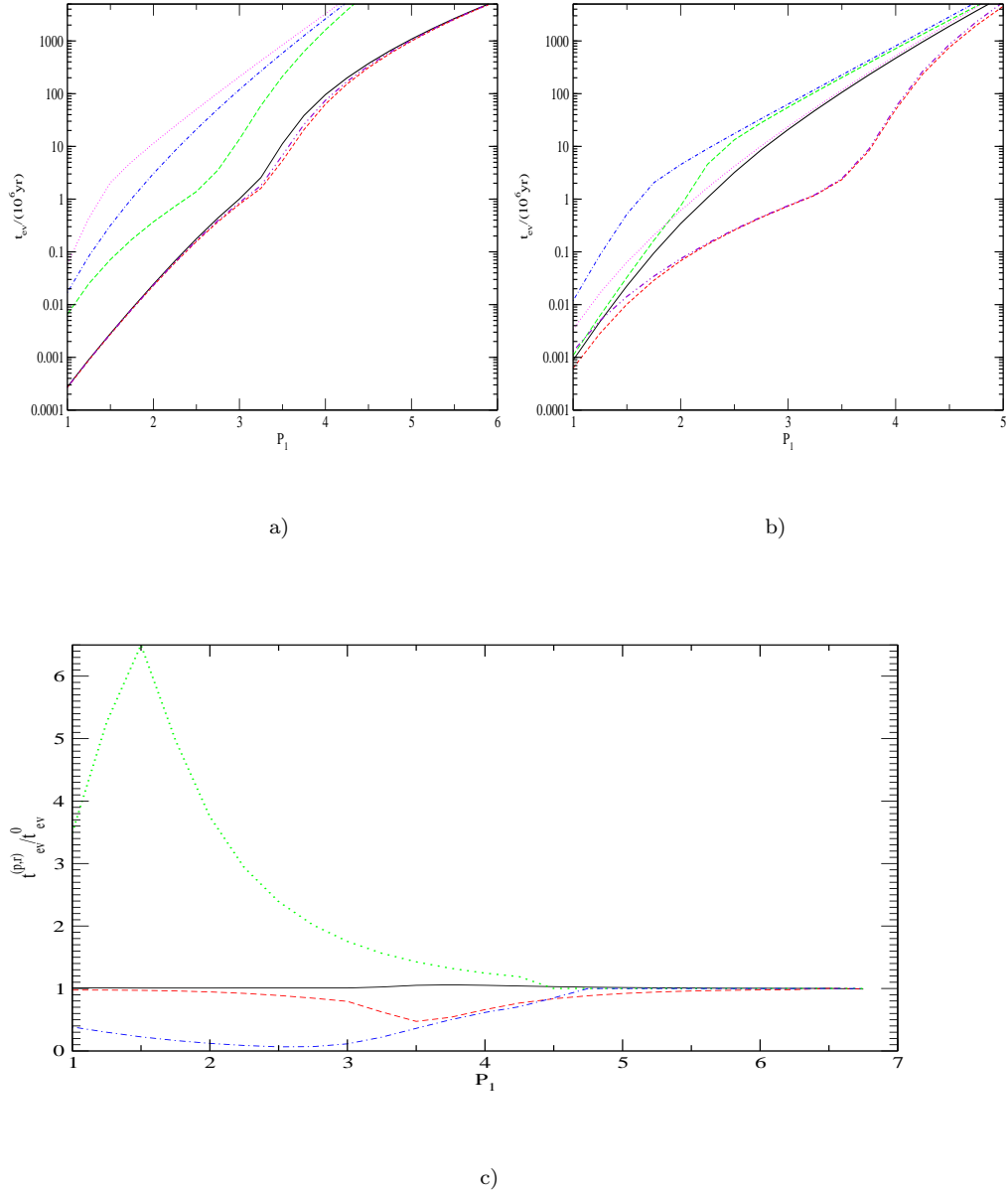

\vspace{1.2cm}
\subfigure[\empty]{\label{fign1}
  \includegraphics[width=0.55\textwidth,height=6cm]{tev.eps}}
%\vspace{2cm}
\subfigure[\empty]{\label{fign2}
\includegraphics[width=0.55\textwidth,height=6cm]{tev5.eps}}
% \vspace{2cm}

 \begin{minipage}[b]{1.0\linewidth}\centering
 \vspace{0.5cm}
\hspace{2cm} { a)} \hspace{6cm} { b)}
\vspace{1cm} 
\end{minipage}
 %\vspace{2cm}

%\subfigure[\empty]{\label{fign2}
%\includegraphics[width=\textwidth,height=7cm]{tev5.eps}}

\subfigure[\empty]{\label{fign1n}
 \includegraphics[width=\textwidth,height=6cm]{r.eps}}

%\hspace{10cm}
%\vspace{ 2cm}
%\\
%\hspace{5cm} a)\hspace{5cm} b)
% \vspace{1cm}

\begin{minipage}[b]{1.1\linewidth}\centering
\vspace{0.5cm} 
\hspace{16cm}{ c)}
\vspace{0.5cm} 
\end{minipage}

% \vspace{1cm}

%\end{figure}
%\begin{figure}

\caption{a) The evolution time scale $t_{ev}$ as a function of 
 $P_1$ in units of 1 day.
The planet's mass $M=1M_{J}.$ Curves are numbered  below in order of increasing magnitude at a fixed value of
$P_1$ for $P_1 > 2.$
The solid curve (3) corresponds to $t_{ev}$  
obtained with help of (\ref{eqn7}) together  with $\Delta E$ calculated for a non-rotating star.
The short dashed curve (1) is   for the retrograde 
  case with  $\Omega (t)$  given by (\ref{eqn1}).
The corresponding plot for the  prograde  case almost coincides with the solid curve, and 
  so is not shown.
The long dashed (4), dot dashed (5) and dotted curves (6)  are  for  tides  excited in the star 
$(\Delta E_{pl}=0)$. They are for the   retrograde, a non-rotating and   prograde cases, respectively. The dot dot
dashed curve (2) is  for the retrograde case with all tides 
 , but with the energy transfer
 $\Delta E$ reduced by a factor of 1.5 as  compared to the value  given
by (\ref{e28}).
b)  As in a) but for $M=5M_J$ and with the difference that  in this case
 the long dashed (5) , dotted (4) and dot dashed curves (6) 
determined  from tides exerted in the star alone  correspond
to the  retrograde,
non-rotating  and  prograde cases respectively.
c) The ratios of the evolution time
 scales corresponding to the prograde and retrograde cases,
 $t_{ev}^{p}$ and $t_{ev}^{r},$ to the time scale corresponding to
 the non-rotating star, $t_{ev}^{0}$, are plotted as functions of
 $P_1$. The solid and dashed curves respectively apply when 
tides exerted in the star and  planet 
are both included.  The dotted and dot
 dashed curves respectively apply when only tides in the star are included.
}
\end{figure}

It is well known that the mode energy, $E_m$, is related to the mode angular momentum, 
$J_m$, as $E_m=\omega_m J_m/m$, see e.g. Friedman $\&$ Schutz (1977). 
Accordingly, the  total amount of energy and angular momentum, $E_t$ and
$J_t$, transferred from the orbit 
to the star obeys a similar relation, where, in general, we should sum 
over all modes of the star. In our case of a slowly rotating star only 
$m=2$ $g$-modes  of  sufficiently  high order
give a significant contribution to the sum.  For our problem the
corresponding eigenfrequencies of these modes can be roughly estimated as
$\omega_n \approx 0.5 \Omega_*$, and we obtain $E_t \approx \Omega_* J_t/4.$  
After the tidal circularisation has happened assuming that the tides
induced in the star dominate over those in the planet we can equate the
total mode energy $E_t$ to the orbital binding energy of the planet
after the circularisation, $E_f={GMM_*/(2 a_f)}$, where $a_f$ is the 
final value of the orbital semi-major axis, and we suppose that $e\sim
0$, and, accordingly, the orbital angular momentum
$J_f=M\sqrt{GM_*a_f}$. From these expressions we obtain
\begin{equation}
{J_t\over J_f}\approx 2\left({R_{*}\over a_f }\right)^{3/2}={1\over \sqrt 2\eta}, \label{eqnn1}
\end{equation}   
where in the last equality we use equation (\ref{e27}) setting $q=0$ there, and assuming that
$J_t/J_f$ is small, and, therefore, the orbital angular momentum is
approximately conserved during the tidal circularisation. In this case we can set $R_{min}=a_f/2$ in
(\ref{e27}). As follows from equation (\ref{eqn2}) below, $J_t/J_f <
0.1$ for the Hot Jupiters with the orbital periods $> 2.4$ days. As seen
from Figs. \ref{fign1} and \ref{fign2} the stellar tides can dominate for
even larger periods, 
accordingly the assumption of approximate conservation of the orbital angular momentum
can then  be justified\footnote{When the planetary tides
dominate,   equation (\ref{eqnn1}) clearly overestimates $J_t/J_f$.}.

This fact also
allows us to link a value of $\eta $ to a value of 
orbital period after the tidal circularisation process has been completed and the planet's orbit has became 
almost circular, $P_{obs}=2\pi\sqrt{{a_f^3/( GM_*)}}$, as
\begin{equation}
\eta \approx 3\left({M_*\over M_{\odot}}\right)^{1/2}\left({R_{\odot}\over R_{*}}\right)^{3/2}P_{1}, \label{eqn2}
\end{equation}     
where $P_{1}=P_{obs}/1day$. This equation relates an observable quantity, $P_{obs}$,
 to the quantity determining the strength of 
the tidal interactions. 

For simplicity, it is assumed below in this section 
that we can simply add the values of energy transfer during successive periastron 
passages either due to sufficiently efficient dissipation of energy stored in the modes between
periastron  passages (e.g. Kumar $\&$ Goodman 1996) or due to the operation
of  stochastic instability in the system, e.g. Kochanek (1992), Kosovichev $\&$ Novikov (1992), Mardling (1995)a,b,  
Ivanov $\&$ Papaloizou (2004), (2007).
 Note that in the latter case the characteristic time scales for  orbital circularisation 
must be understood in some average sense. Some estimates relevant to a more realistic situation are discussed 
in the next section. 

Under this assumption Ivanov $\&$ Papaloizou (2004), (2007)  derived an equation determining 
the rate of change of the orbital semimajor axis, $a,$  in  the form

\begin{equation}
{\dot a_{10}\over a_{10}}=-{1\over t_{10}(t)\sqrt{a_{10}}}, \label{eqn3}
\end{equation}   
where $a_{10}=a/(10au)$ and
\begin{equation}
t_{10}=7\cdot 10^{8}\left(\sqrt{\frac{M_{*}}{M_{\odot}}} 
\frac{ M_{J}}{ M}\right)\left({R_{pl}\over R_{J}}\right)\left({10^{-9}\over \epsilon_{DT}}\right)yr, \label{eqn4}
\end{equation}     
where $M_J$, $R_{J}$ are the Jupiter mass and radius, respectively, $R_{pl}$ is the planet's radius, and
\begin{equation}
\epsilon_{DT}=(\Delta E + \Delta E_{pl})\left({R_{pl}\over GM^{2}}\right), \label{eqn5}
\end{equation}    
where $\Delta E$ is given by equation (\ref{e28}) and $\Delta E_{pl}$ is the energy transfer to the modes exited 
in the planet\footnote{Note that in Ivanov $\&$ Papaloizou (2004),
(2007) there is a misprint in their equations defining
$t_{10}$. Namely, the value of $t_{10}$ is proportional to
${\frac{M_{*}}{M_{\odot}}}$ there instead of the correct square root
dependence of this ratio as in (\ref{eqn4}). However, in these
papers only the case $M_*=M_{\odot}$ was considered and this
misprint has no influence on the results of the papers.} .  
This is calculated in  Ivanov $\&$ Papaloizou (2004),( 2007)
for the value of $\eta$  given by (\ref{eqn2}) corresponding to a specified final orbital period. 
Note that both $\Delta E$ and $\Delta E_{pl}$ depend 
on time. In the case of $\Delta E$ this  is through  the dependence of the stellar angular velocity  on time,  assumed to be
given by equation (\ref{eqn1}).  In the case of $\Delta E_{pl}$ this is mainly through the evolution of the planet's radius  
on time. Following Ivanov $\&$ Papaloizou ( 2004), (2007)  we assume in this paper that this 
evolution is the same as the evolution of an isolated planet in which  the planet's radius shrinks with time. Thus, 
we neglect the possible influence of tidal heating on the planet's structure.

Equation (\ref{eqn5}) can be integrated to give
\begin{equation}
a_{10}(t)=a_{in}\left(1-{1\over 2\sqrt a_{in}}\int^{t}_{t_{in}}{dt'\over t_{10}(t')}\right)^2, \label{eqn6}
\end{equation}
where $a_{in}(t_{in})$ is the 'initial' value of the planet's semimajor axis (in units of $10au$) 
just after  tidal capture has occurred. Let us define the evolution time scale as the time corresponding to 
$a_{10}$ being formally equal to zero. From (\ref{eqn6}) we get
\begin{equation}
\int^{t_{ev}}_{t_{in}}{dt'\over t_{10}(t')}=2\sqrt a_{in}, \label{eqn7}
\end{equation} 
which provides an implicit equation for determination of $t_{ev}.$ However, let us recall that the evolution
time scale so defined describes the evolution of the semi-major axis at large scales, of the order 
of $1-10au$ or larger, under the assumption that the dynamical tides can effectively transfer the orbital energy
to energy associated with normal modes and thermal energy of the star and the planet. At smaller scales 
the dynamic tides are clearly ineffective and other processes (e.g.due to so called quasi-static tides ) must 
be considered to explain final stages of circularisation of exoplanets, see the next section.

The results of the solution of this equation for $a_{in}(t_{in}) =1$ corresponding to an initial orbit
with semi-major axis of $10au$ and a central solar mass  are shown 
in Figs.\ref{fign1}, \ref{fign2}, 
for $M=1M_{J}$ and $M=5M_{J}$, respectively. In both cases it is assumed that the planet is in a state of 
rotation corresponding to  so-called pseudosynchronization, see Ivanov $\&$ Papaloizou (2004), (2007).
 We consider only the case 

of the planet's orbit being coplanar with the stellar equatorial plane but with both retrograde and prograde directions of orbital motion.
As seen from Fig. \ref{fign1} when tides are exerted both in the star and in the planet,  there is a difference in the circularisation
time scale corresponding to the case of  retrograde motion as compared to the cases  of either a non-rotating star or of  prograde motion. 
Since criteria for the  dissipation of  the energy in normal modes
 as well as conditions determining onset of the stochastic instability may be different
for the planet and for the star and it may happen that only stellar tides are important for the circularisation problem under certain 
circumstances, we also plot 
the curves calculated  assuming that $\Delta E_{pl}=0$. As seen from Fig. \ref{fign1} the stellar tides themselves can provide 
a tidal circularisation time scale smaller than a typical life time of
a planetary system of  $\sim 5\cdot 10^9yr$ for $P_{orb} < 4$ days.
In Fig. \ref{fign1n} we show the ratios of the evolution time scale
corresponding to the prograde and retrograde cases to the case of
a non-rotating star, for a planet with $M=1M_{J}$. One can see from this
figure  that assuming that  tides in the star and in the planet are both
effective,  there is practically no difference between the prograde
and non-rotating cases,  while the retrograde case has an evolution
time scale $1.2-2$ smaller,  for final  orbital periods
in the range  3.5 to 4.5 days.  
When only the stellar tides are taken into account,  the difference
between the prograde, retrograde and non-rotating cases can be quite
significant for smaller final  orbital periods.   
%\begin{figure}
%\vspace{1.5cm}
%\includegraphics[width=\textwidth]{tev5.eps}
%\vspace{1cm}
%\caption{Same as Fig (\ref{fign1}) but for $M=5M_J$. The solid, short dashed and 
%long dashed curves are calculated assuming that the
%tides are exerted both in the planet and in the star, for the
%non-rotating star, the case of retrograde rotation and  the case of
%prograde rotation, respectively. The dot dot dashed, dotted and dot dashed curves are 
%determined only from tides exerted in the star,  and correspond
%to  retrograde rotation, 
%a non-rotating star and  prograde rotation respectively.}
%\label{fign2}       % Give a unique label
%\end{figure}
Since $\Delta E$ grows with the mass of the planet the contribution of 
stellar tides is more important for the $M=5M_J$ case shown in Fig. \ref{fign2}. 
In this case $t_{ev}$ is practically determined by these tides 
for $P_{orb}$ larger than 2-2.5 days and the difference between 
the retrograde, prograde and non-rotating case is quite prominent.    

As  mentioned above there is a difference between the numerically and analytically obtained values of $\Delta E$, which 
is more significant for a sufficiently rapidly  rotating star and  retrograde orbits. Let us estimate the possible 
difference in the evolution time scale  associated with this assuming that $t_{ev} > 10^7yr$. From equation  (\ref{eqn1}) it follows that 
$\Omega (t > 10^{7}yr) < 0.08\Omega_*$, and from Figs. \ref{fign1} and \ref{fign2} it follows that when $\Omega /\Omega_{*} 
< 0.08$ the ratio of analytically obtained values of $\Delta E$ to the numerical ones is of the order of or smaller than
1.5. In order to see the effect of  tides exerted in the star being reduced by a factor 1.5,  we plot 
the additional dot dot dashed curve in Fig. \ref{fign1} corresponding to the retrograde case with the energy transfer given by
equation (\ref{e28}) reduced  by a factor of  1.5.
 It is seen that the curve is very close to the short dashed curve calculated
with help of (\ref{e28}). Therefore, it seems that
 this issue is not significant for a one Jupiter mass planet. It is 
even less important for more massive planets whose evolution time
 scales show a stronger dependence on the relative direction 
of the orbital motion and of the stellar rotation, see e.g. Fig. \ref{fign2}.

\subsubsection{Conditions leading to stochastic instability or effective dissipation of the mode energy}\label{KGS}

%As discussed in  Kochanek (1992), Kosovichev $\&$ Novikov (1992), Mardling (1995)a,b, Mardling $\&$ Aarseth 2001, 
%Ivanov $\&$ Papaloizou (2004), (2007),  when the orbital semi-major axis is larger than some critical value,
%$a_{st}$, stellar and planetary perturbations excited at successive periastron passages have approximately uncorrelated phases. 
%This happens when a  phase change of perturbed quantities associated with a mode 
%during one orbital period determined by a change of the orbital period is larger than 
%some critical number of order of unity. The change of orbital period is, in turn, 
%due to the energy transfer from orbit 
%to the modes or back.
% This number was numerically estimated by Ivanov $\&$ Papaloizou
%(2004) in the simplest case of only one mode interacting with orbital motion due to tides. 

 As discussed in  Kochanek (1992), Kosovichev $\&$ Novikov (1992),
Mardling (1995)a,b, Mardling $\&$ Aarseth 2001,
Ivanov $\&$ Papaloizou (2004), (2007),  when the orbital
semi-major axis is larger than some critical value,
$a_{st}$, stellar and planetary perturbations excited at successive
periastron passages have approximately uncorrelated phases.
This happens when the  phase change of perturbed quantities associated with a normal mode
occurring during one orbital period, that arises on account of the change induced in 
 the orbital period itself, is larger than 
some critical number of order of unity. This change in the orbital period  results from
 the energy interchange between the orbit       
and  the normal modes. This number was estimated numerically by Ivanov $\&$ Papaloizou
(2004) in the simplest case of only one mode interacting with  the orbital motion due to tides.

In the situation where the phases are uncorrelated the mode energy, which is proportional to 
square of the mode amplitude grows on average  in  proportion  to  the  number of periastron passages. This
effect is referred to as the stochastic instability. 

\begin{figure}
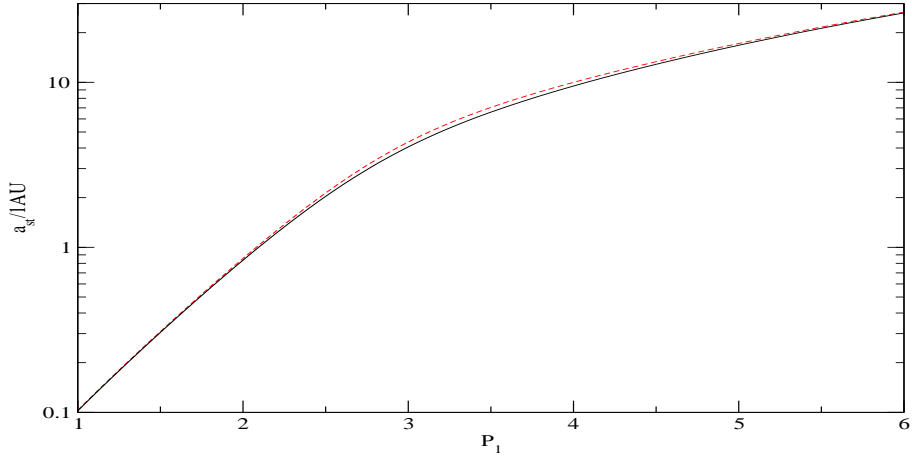

%\begin{center}$
%\begin{array}{c}
%\\
%\\
%\\
 \vspace{1cm}
\subfigure[\empty]{\label{nfig1}
\includegraphics[height=6cm,width=\textwidth]{ast.eps}}
%\vspace{1.5cm}

\begin{minipage}[b]{1.1\linewidth}\centering
\vspace{1cm}
 \hspace{0.1cm} { a)}
\vspace{1.5cm}
 \end{minipage}
\subfigure[\empty]{\label{nfig1b}
\includegraphics[height=6cm,width=\textwidth]{ast_107.eps}}   
\begin{minipage}[b]{1.1\linewidth}\centering
\vspace{1cm}
 \hspace{0.1cm} { b)}
\vspace{1.5cm}
 \end{minipage}

%&\hspace{0.5cm}&
%\\
%\\
%\\
%\end{array}$
%\end{center}
\caption{ a)
 The scale $a_{st}$ as a function of  time $t$. The rotational history of the star is given by
(\ref{eqn1}). Solid curves correspond to the retrograde orbits with 
 values of $P_1=1,2,3,4,5$ with curves having larger values  at given moment of time 
corresponding to larger values of $P_1.$  The difference between the prograde and retrograde orbits is not well pronounced.
As an example the dashed curve shows the dependence of $a_{st}$ on $t$ 
for the prograde orbit with $P_1=5$.
b)  Same as the  upper
panel but $a_{st}$ is shown as a function of $P_1$ for a time 
or stellar age of  $t=10^7yr$. The solid and dashed curves show the retrograde and prograde cases,
respectively.}
%\label{nfig1}
\end{figure}

From the results obtained in  Ivanov $\&$ Papaloizou (2004) it follows that the stochastic instability sets in when
\begin{equation}
a > a_{st}=(\tilde \omega \epsilon_{DT})^{-2/5}\left({M_J\over M}\right)^{3/5}\left({R_{pl}\over R_J}\right)au, 
\label{ip2004}
\end{equation} 
where $\tilde \omega =\omega /\Omega_*$, $R_{pl}$ is the planet's radius and $R_J$ and $M_J$ are the radius and
mass of Jupiter, respectively. We use equation (\ref{ip2004}) to estimate
$a_{st},$  for definiteness setting  $\tilde \omega =0.5$ there, which is appropriate for $g$ modes excited in the star
and $m_{pl}=M_{J}$, $R_{pl}=R_{J}$. Since both stellar and planetary tides depend on time $t$ 
we plot the dependence of $a_{st}$ on $t$ in Fig. \ref{nfig1} for different values of $P_1=1,2,3,4$ and $5$.

%\vspace{1cm}
%\begin{figure}[h]
%\begin{center}$
%\begin{array}{ccc}
%& & \\
%& & \\
%& & \\
%\includegraphics[height=2.8in,width=2.2in]{ast.eps}
%&\hspace{0.5cm}&
%\includegraphics[height=2.8in,width=2.2in]{ast_107.eps}
%\end{array}$
%\end{center}
%\caption{Left panel:
% The scale $a_{st}$ as a function of  time $t$. The rotational history of the star is given by
%(\ref{eqn1}). Solid curves correspond to the retrograde orbits with
% values of $P_1=1,2,3,4,5$ with curves having larger values  at given moment of time
%corresponding to larger values of $P_1.$ 
% The difference between the prograde and retrograde orbits is not well pronounced.
%As an example the dashed curve shows the dependence of $a_{st}$ on $t$
% for the prograde orbit with $P_1=5$. Right panel: Same as the upper
%panel but $a_{st}$ is shown as a function of $P_1$   for a time or stellar age
%of  $t=10^7yr$.
% The solid and dashed curves show the retrograde and prograde cases,
%respectively.}
%\label{nfig1}
%\end{figure}

%\vspace{1cm}
%\begin{figure}
%\includegraphics[width=\textwidth]{ast.eps}
%\vspace{1cm}
%\caption{The scale $a_{st}$ as a function of  time $t$. Solid curves correspond to the retrograde orbits with 
% values of $P_1=1,2,3,4,5$ with curves having larger values  at given moment of time 
%corresponding to larger values of $P_1.$  The difference between the prograde and retrograde orbits is not well pronounced.
% As as example the dashed curve shows the dependence of $a_{st}$ on $t$ for the prograde orbit with $P_1=5$.}
%\label{nfig1}       % Give a unique label
%\end{figure}
  
One can see from this figure  that the critical semi-major axis $a_{st}$ is typically of the order $0.1-10au.$  Note, however,
that  the analysis of Ivanov $\&$ Papaloizou 2004 can  somewhat overestimate the value of $a_{st}$ since it was done under the 
assumption that only one normal  mode plays a significant role.  For  the problem on hand,  several modes
are excited in the star and and in the  planet which could give a comparable contribution to the energy exchanged  with the 
orbit. This is likely to increase 'the  degree of stochasticity'  in the system leading to  smaller 
values of $a_{st}.$ However, in any case it seems highly improbable that dynamic tides can operate in the stochastic regime
at scales much smaller than, say, $0.1au,$, for the whole of the  interesting range of $P_1.$  Thus,  provided that dissipation of 
 mode energy is not unexpectedly large,   dynamical  tides are not efficient at such scales so that  quasi-static tides need 
be invoked to explain further tidal circularisation.

%\begin{figure}[h]
%\begin{center}$
%\begin{array}{ccc}
%& & \\
%& & \\
%& & \\
%\includegraphics[height=2.8in, width=2.2in]{tnl.eps} &\hspace{0.5cm}&
%\includegraphics[height=2.8in, width=2.2in]{rat.eps}
%\end{array}$
%\end{center}
%\caption{Left panel: 
%The characteristic time of non-linear dissipation, $t_{nl}$ as a function 
%of the dimensionless orbital period 
%after circularisation, $P_1$ for a star with rotational period 1 day. 
%The solid and dashed curves correspond to the prograde and retrograde cases, respectively. Right panel:
%Ratio of the energy transfer $\Delta E$ to the threshold energy $E_{thr}$
% for the parametric instability to 
%operate. The value of $E_{thr}$ is taken to be $10^{35}ergs$, see the text for discussion.}
%\label{nfig2}
%\end{figure}

\begin{figure}
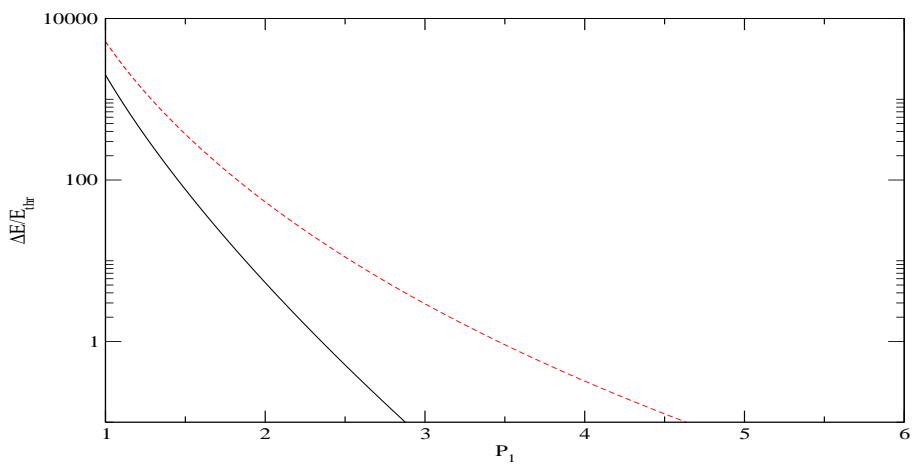

%\begin{center}$
%\begin{array}{c}
% \\
% \\
% \\
 \vspace{1cm}
\subfigure[\empty]{\label{nfig2}
\includegraphics[height=6cm,width=\textwidth]{tnl.eps}}
%\vspace{1.5cm}

\begin{minipage}[b]{1.1\linewidth}\centering
\vspace{1cm}
 \hspace{0.1cm} { a)}
\vspace{1cm}
 \end{minipage}
%&\hspace{0.5cm}&
%\\
%\\
%\\
\subfigure[\empty]{\label{nfig2b}
\includegraphics[height=6cm,width=\textwidth]{rat.eps}}
%\vspace{1.5cm}

\begin{minipage}[b]{1.1\linewidth}\centering
\vspace{1cm}
 \hspace{0.1cm} { b)}
\vspace{1cm}
 \end{minipage}
%\end{array}$
%\end{center}
\caption{ a)
The characteristic time of non-linear dissipation, $t_{nl}$ as a function of the dimensionless orbital period
after circularisation, $P_1$ for a star with rotational period 1 day.
The solid and dashed curves correspond to the prograde
 and retrograde cases, respectively.
b) Ratio of the energy transfer $\Delta E$ to the threshold energy $E_{thr}$
 for the parametric instability to
operate. The value of $E_{thr}$ is taken to be $10^{35}ergs$, see the text for discussion.}
%\label{nfig2}
\end{figure}

%\begin{figure}
%\includegraphics[width=\textwidth]{tnl.eps}
%\vspace{1cm}
%\caption{The characteristic time of non-linear dissipation, $t_{nl}$ as a function of the dimensionless orbital period 
%after circularisation, $P_1$ for a star with rotational period 1 day. 
%The solid and dashed curves correspond to the prograde and retrograde cases, respectively.}
%\label{nfig2}       % Give a unique label
%\end{figure}

\begin{figure}
\vspace{1.5cm}
\includegraphics[width=\textwidth]{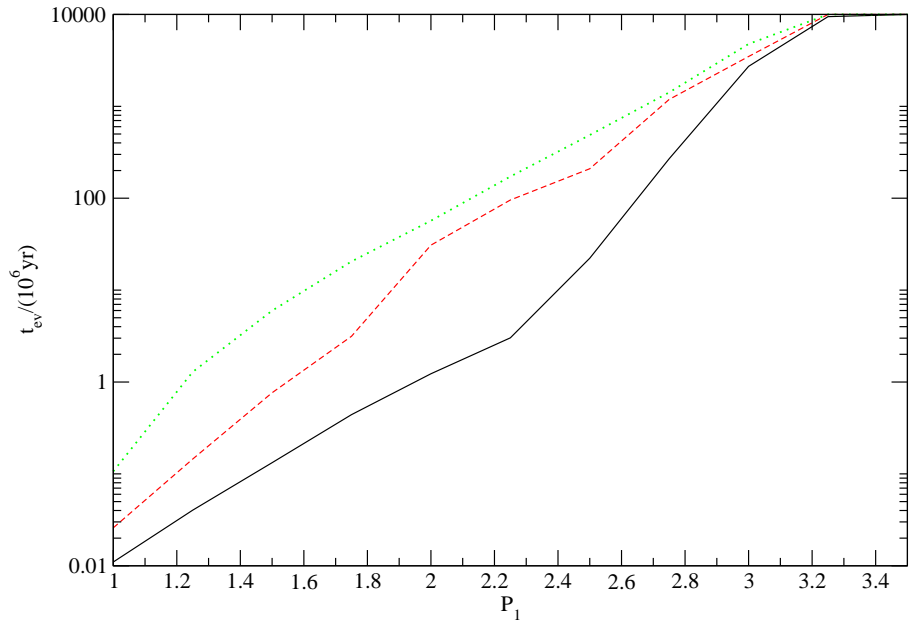}
\vspace{1cm}  
\caption{The evolution time-scale $t_{ev}$ calculated under the  assumption that only  tides exerted in the star 
are effective  and that the energy transfer is proportional to the attenuation factor given by (\ref{neqq2}). Note that 
here we define $t_{ev}$ as a time-scale of evolution of the semi-major axis from its initial value equal taken to be 
$10au$ to a final value taken to be $0.5au.$  The solid, dashed and dotted curves correspond to retrograde, 
non-rotating and prograde cases, respectively. The planet's mass $M=1M_{J}$.}
\label{nfig3}       % Give a unique label
\end{figure}  

However, in addition to the stochastic instability there is another potentially important process, which may facilitate 
the orbital evolution at large values of $a.$    This is through the so-called non-linear dissipation of the  energy $E_{nm}$ stored 
in the perturbed g-modes of the star (Kumar $\&$ Goodman 1996, hereafter KG). This results from the  parametric excitation of 
low frequency high order normal modes through their non-linear interaction  with the modes excited by tides. 
Provided that this energy is larger than a certain threshold value, $E_{thr}$, the characteristic time scale 
of dissipation of the energy, $t_{nl}$ is given by the expression (KG)
\begin{equation}
t_{nl}\approx  10\sqrt{ \left({10^{38}ergs\over E_{nm}}\right ) }yr. 
\label{neqq1}
\end{equation}    
When this time is smaller than than the orbital period $P_{orb}$ the energy transferred during a single periastron 
flyby to the normal model can be dissipated in one orbital period, and, accordingly,  the expression for $\Delta E$ 
can be simply added to describe the effect of successive periastron passages.
 On the other hand, when $f=P_{orb}/t_{nl} \ll 1,$
assuming there is no build up, the energy transferred from the orbit to the star that is actually dissipated
 should be attenuated by a factor of the order of $f$. A more accurate
calculation, which takes into account a possibility of resonances between the orbit and a mode,  and
therefore a build up transferred energy in the star,  gives  what we describe as an attenuation 
factor, $f_{a}$, in the form (see Appendix C)
\begin{equation}
f_{a}=(1-\exp (-2f))(1+\exp (-2f)-2\exp (-f)\cos (2\pi \omega_{n}/\Omega_{orb}))^{-1},
\label{neqq2}
\end{equation}   
which is valid for  $0 < f < \infty $. Here $\omega_{n}$ is a typical value of eigen frequency of a mode
excited by tides. For our problem it is of the order of $0.5\Omega_{*}.$  Note that $f_{a}\rightarrow 1$ in the limit
$f\rightarrow \infty $ as expected. In order to allow for  the   weakening of the orbital energy transfer 
in the limit $f\rightarrow 0$  we adopt  $f_a\Delta E$ for the energy exchanged per pericentre 
passage in equation (\ref{eqn5}) instead of $\Delta E$ itself.  

We show the dependence of $t_{nl}$ on $P_1$ in Fig. \ref{nfig2} for a star with rotational period 1 day
and the evolution time scales calculated  taking 
into account the attenuation factor (\ref{neqq2}) are shown  in Fig. \ref{nfig3}. 
As seen from Fig. \ref{nfig2} there is a marked difference between
the cases of prograde and retrograde motion with the retrograde case corresponding to much smaller values of $t_{nl}$ at
given value of $P_{1}$. This can increase the difference between probabilities of tidal capture of planets moving
in opposite directions with respect to the stellar rotation.  It is also important to note that the threshold value of 
the energy $,E_{thr},$  for our case can be estimated to be  $E_{thr}\sim 10^{35}ergs.$ for a star with solar values of
mass and radius and rotational period of order of a few days\footnote {In this case KG estimate the value of $E_{thr}$ 
to be approximately $10^{48}(\omega t_{th})^{-6/5}ergs$, where $\omega $ is a typical eigen frequency of a mode 
undergoing parametric instability and $t_{th}$ is the stellar thermal time. Taking $\omega \approx 0.5\Omega_{*}\sim
3\cdot 10^{-4}s^{-1}$ and $t_{th}\sim 3\times 10^{7}yr$ we get $E_{thr}\sim 10^{35}ergs$.} At sufficiently large values of $P_1$ (see below) 
the amount of energy transferred during one periastron flyby of a planet moving on a prograde orbit is significantly
smaller than $,E_{thr},$  making the non-linear dissipation of the mode energy impossible. This 
can further amplify the difference between the capture probabilities.  Interestingly, the values of $t_{nl}$ for 
retrograde orbits are of the order of the orbital periods of planets with semimajor axes of order  $10au$ - the
scale, where the processes of tidal capture are expected to operate. 

In order to calculate $t_{ev,}$ as shown in Fig. \ref{nfig3},  for simplicity we set  $\Delta E_{pl}=0$ in (\ref{eqn5}). 
One can see from this figure  that when the attenuation effect is taken into account,  dynamic tides can explain the initial stages of 
tidal circularisation up to $P_{1} < 3$ in contrast to the simpler case considered in the previous section, 
where the beginning of effective  circularisation is expected up to $P_{1} < 4,$  see Fig. \ref{fign1}. We also checked that  
typical values of $\Delta E$ used to calculate $t_{ev}$ are approximately equal to $E_{thr}$ when $P_{1}\sim 3.$ 
Thus, at larger values of $P_1,$  non-linear dissipation of  mode energy is likely to be unimportant.  

The results discussed in this section confirm our previously made assertion that dynamic tides can lead to an 
efficient decrease of semimajor axis on scales corresponding to tidal capture both as a result of the stochastic instability 
and the possibility of non-linear dissipation of energy stored in normal modes excited in the star. 
However, at smaller scales,  the evolution of $a$ due to  the processes we have discussed  clearly stalls,  
necessitating the consideration of different nonlinear effects, such as wave breaking (e.g. Barker $\&$ Ogilvie (2010),
Barker (2011)), operating on normal modes, or other phenomena such as turbulent  friction operating on 
quasi-static tides to be able to describe  later stages of orbital evolution.

\section{Discussion}\label{Discuss}

In this paper we have developed an analytic theory for calculating the tidal response of a slowly rotating star
resulting from a fly by of  a planetary perturber in a parabolic orbit, with orbital plane at an arbitrary inclination
to the stellar equatorial plane. Expressions for the energy transferred to the  star were presented in
section \ref{energytr} (see equations (\ref{e22})-(\ref{e28})). In order to develop
these expressions only the first order corrections to the normal
mode eigenfrequencies appropriate to  the non rotating star were incorporated.
This implicitly assumes that these modes only change slightly and
continue to dominate the tidal response in this form.
In order to confirm these results
numerical solutions of the  linear  tidal problem were considered in \ref{Numerical}.
In these calculations Coriolis forces, but not centrifugal forces, were taken into account.
Thus the equilibrium model was spherical.
Beyond this no assumption of slow rotation was made. 
In practice the calculations were done for a polytrope of index $n=3.$
Detailed results for prograde encounters, retrograde encounters and encounters with orbital plane
perpendicular to the equatorial plane were presented in 
section \ref{Numres}. These were compared with results obtained from the analytic treatment in
section \ref{Numanalcomp}. In general the agreement between the two approaches was good.
Both agreed that the tidal interaction was significantly more effective for retrograde
encounters. The energy exchanged between the orbit and the 
star  was about one  order  of magnitude larger for retrograde rotation with
$\Omega=0.1\Omega_*$ compared to prograde rotation with the same angular velocity when $\eta=4\sqrt{2},$
with the deviation becoming larger as $\eta$ increased.
Nonetheless the numerically obtained results showed an increasing deviation from the
analytically determined ones for increasing angular velocity of retrograde rotation.
The maximum  departure  was about a factor of two for $\Omega_* =0.25$ and $\eta=4\sqrt{2}$
increasing to  a factor of $2.5$ when $\eta$ was increased to $8.$  However,
due to the very rapid decline in the strength of tidal effects with increasing pericentre distance,
we note that such deviations are scarcely significant for practical applications.

The time scale for the initial stage of tidal circularisation inferred from our results 
for an orbit with initial semi-major axis of $10au$ was given in section \ref{circtime}
for planets of $1M_J$ and $5M_J.$ The rotation of the star, taken to be
one solar mass, was assumed to follow the Skumanich (1972) law. 
 When tides from both the planet were included, initial
circularisation down to circular orbits of periods $< 5 days$  could be readily accounted
 for both masses. When the effects of tides in the planet were neglected, our results
indicated that tides acting in the star alone could be capable 
of producing a significant change in the planet's semi-major axis 
for final periods $< 4 days$ for both masses. Provided that some additional processes 
such as quasi-static tides are sufficiently efficient at smaller scales our time scales may give 
an estimate of a characteristic time of the whole stage of circularisation.  
\subsection{Application to the circularisation of Hot Jupiters}
The discovery of Hot Jupiters in orbits highly inclined to the stellar equatorial
plane with some orbits even being retrograde has given  support to the idea
that at least some of these objects were scattered into highly elongated orbits that were
the circularised by tidal effects. The process of the excitation
of oscillations by   tides acting  on the planets and the consequent circularisation 
has been discussed by Ivanov \& Papaloizou (2004, 2007, 2010) and Papaloizou\& Ivanov (2010).
Here this work has been extended to consider tides excited in the star.
These may be significant because in contrast to tides acting on the planet,
they are not equally effective for prograde and retrograde rotation.
This feature should be borne in mind when considering the distribution of objects as a function
of inclination. Hebrard et al. (2010) point out that of the $37$ systems for which 
the inclination between the orbital and stellar equatorial planes has been measured at the time of writing,
about one third are strongly misaligned with the distribution of inclinations being dependent
on the planet mass such that retrograde systems only appear for $M < \sim 2.5M_J.$
Our results indicate that under the assumptions we  made about the stellar angular velocity,
circularisation times for prograde and retrograde orbits should differ
by a factor of two at least initially, for a 'typical' period
after circularisation of the order of four days, 
and that systems should be seen up to the same limiting period in both types of system.
There would thus be some asymmetry produced by the effect of tides in
this case though this claim should be confirmed by numerical N-body
experiments  incorporating  planet-planet scattering together with tidal effects
that take account  of  effects  resulting from  stellar rotation.
The asymmetry may be more pronounced for masses $M> \sim 5M_J.$ However, at present there are relatively
few objects with measured orbital inclinations with respect to the stellar rotation axis
measured in this mass range. Future observations may clarify this issue.
Similarly results may differ for different histories
of the stellar rotation frequency and/or different stellar masses. 
 Winn et all (2010) point out that there is an increase in  the
fraction of  planets on inclined orbits for stars with larger
masses. These may have been rotating  more rapidly in the past,  thus increasing the
asymmetry between the prograde and retrograde tidal interactions.  

It is clear that the result for the energy transfer obtained in this
paper can also be directly applied to a general situation where two rotating
stars of comparable mass with misaligned directions of rotational
axes tidally interact with each other. Therefore, our results may be 
of interest for the problem of tidal capture of stars in young stellar
clusters.

Our results can also be generalised in several ways. 
Firstly, expressions for the transfer of angular momentum can be obtained
by a calculation similar to what is made in this paper for the energy
transfer. Note that all three components of the angular momentum
( with respect to the coordinate system defined with respect to
the stellar rotation axis) are transferred in the case of a general
inclined encounter. Secondly, a more realistic  stellar model   should be
used to calculate the eigenfrequencies, overlap integrals and the
frequency splitting coefficients $\beta_r$. Thirdly, in order 
to reach  better agreement between the theory and the numerical
calculations,  a better theory of the  tidal excitation of high order
$g$-modes in a rotating star should, in principle, be developed. In the
framework of such a theory one may try to relax the assumption of
smallness of the stellar angular velocity   in comparison to the mode eigenfrequencies. 
These issues will be addressed  in  future investigations.

%as required. Don't forget to give each section
%and subsection a unique label (see Sect.~\ref{sec:1}).
%\paragraph{Paragraph headings} Use paragraph headings as needed.
%\begin{equation}
%\end{equation}

%\begin{figure}
%  \includegraphics{example.eps}
%\end{figure}
%\begin{figure*}
%  \includegraphics[width=0.75\textwidth]{example.eps}
%\end{figure*}
%\begin{table}
%\begin{tabular}{lll}
%\hline\noalign{\smallskip}
%\end{tabular}
%\end{table}

\begin{acknowledgements}
We are grateful to M. Nagasawa and A. V. Tutukov for useful remarks.
This work was supported by the Science and Technology Facilities
Council  [grant number ST/G002584/1], by the Dynasty Foundation,
by the programme ''Research and Research/Teaching staff 
of Innovative Russia'' for
2009 - 2013 years (State Contract No. P 1336 on 2 September 2009)
and by RFBR grant 11-02-00244-a..
PBI thanks DAMTP, University of Cambridge for hospitality. 

This paper was started when both PBI and JCBP took part in the Isaac Newton 
Institute programme 'Disks and Planets'.

\end{acknowledgements}


\begin{appendix}
\section*{ Appendix A}
\section*{Explicit expressions for $d^{(2)}_{m,n}$}
We list below the explicit expressions for the Wigner
$d-$functions, which are used in our calculations. We need
\begin{equation}
d^{(2)}_{2,\pm 2}={(1\pm \cos \beta)^2\over 4},\quad
d^{(2)}_{1,\pm 2}=\pm {\sin \beta (1\pm \cos \beta )\over 2},
\quad d^{(2)}_{0,\pm 2}=d^{(2)}_{\pm2,0}=\sqrt {{3\over 8}}\sin^{2}\beta,
\label{a1}
\end{equation}
and the functions obtained from (\ref{a1}) by the rule
$d^{(2)}_{-n,\pm 2}=(-1)^{n}d^{(2)}_{n,\mp 2}$ for $n=0,1,2$.
Additionally, we use
\begin{equation}
d^{(2)}_{0,0}={3\cos^2 \beta -1\over 2},\quad d^{(2)}_{\pm
1,0}=\mp \sqrt {{3\over 2}} \sin \beta \cos \beta .
\label{a2}
\end{equation}

\section*{Appendix B}
\section*{Transfer of angular momentum during the periastron flyby}
The rate of angular momentum transfer, $\dot {\bf J} $, as well as 
its value of the periastron flyby, $\Delta {\bf
J}=\int^{+\infty}_{-\infty}dt \dot {\bf J}$
are determined by the expression
\begin{equation}
\dot {\bf J}=\int d^3x\rho^{'}({\bf r}\times \nabla U),  
\label{b1}
\end{equation}
where ${\bf r}$ is the radius vector, and $\rho^{'}$ is the density 
perturbation, which can be found from the continuity equation
\begin{equation}
\rho^{'}=-\nabla \cdot (\rho {\bf \xi})=-\sum_{n} {b_{n}(t)\over
  r}({1\over r}{d\over dr}(\rho r^{2}\xi_{r})-6\rho \xi_{s})Y_{2,n},
\label{b2}
\end{equation} 
where we use equations (15), (16) and the known properties of the
spherical functions. Now we substitute equations (3) and (\ref{b2}) in
(\ref{b1}), integrate by part the term proportional to ${d\over
dr}(\rho r^{2}\xi_{r})$ and use (20) to obtain
\begin{equation}
\dot {\bf J}=Q\sum_{n,k} A_{n}b_{k}{\bf J}_{k,n},\quad  {\bf J}_{k,n}=\int
d\Omega Y_{2,k}({\bf e}_{r}\times r\nabla Y_{2,n}), 
\label{b3}
\end{equation} 
where integration is performed over the solid angle, and ${\bf e}_r$ 
in the unit vector in the radial direction. The quantity ${\bf
J}_{k,n}$ can be easily evaluated with help of the theory of the 
vector spherical functions (see e.g. Varshalovich et al 1989) with the
result
\begin{equation}
{\bf J}_{k,n}=i(-1)^{k}[n\delta_{k,-n}{\bf e}_{z}
 (\sqrt{(2-k)(3+k)}\delta_{k,-(n+1)}{\bf e}_{-1}-
\sqrt{(2+k)(3-k)}\delta_{k,1-n}{\bf e}_{1})],
\label{b4}
\end{equation}   
where ${\bf e}_{\pm 1}=\mp {1\over 2}({\bf e}_{x}\pm i{\bf e}_{y})$, and
$\bf{ e}_{x}$, $\bf{ e}_{y}$, $\bf{ e}_{z}$ are the usual unit vectors
associated with the Cartesian coordinates $(x,y,z)$.

In order to calculate $\Delta {\bf J}$ we express the quantities $b_k$
and $A_n$ in (\ref{b3}) through their respective Fourier transforms and
integrate the result over time using (7) to get
\begin{equation}
\Delta {\bf J}=Q\sum_{n,k} M_{n,k}{\bf J}_{k,n}, \quad M_{n,k}=
2\pi (-1)^n\int d\sigma \tilde b_{k}(\sigma) (\tilde A_{-n}(\sigma))^{*}. 
\label{b5}
\end{equation}
Now we substitute (21) in (\ref{b5}) and calculate the integral over
$\sigma $ in (\ref{b5}) taking into account only the resonance
contribution. The calculation is analogous to the one made for the
energy transfer in Section 3 with the result
\begin{equation}
M_{n,k}=i{\pi^{2}\over \omega}Q[(-1)^n\tilde A_{k}(\omega^{+}_k)\tilde
A_{-n} (\omega^{+}_k)^{*}-(-1)^k\tilde A_{n}(\omega^{+}_{-k})\tilde
A_{-k}(\omega^{+}_{-k})^{*}].
\label{b6}
\end{equation}   
Now we substitute (\ref{b4}) and (\ref{b6}) in (\ref{b5}) and get the
angular momentum transfer in the form
\begin{equation}
\Delta {\bf J}=\Delta J^z{\bf e}_{z}+\Delta J^{-1}{\bf e}_{-1}+\Delta
J_{1}{\bf e}^{1},
\label{b7}
\end{equation}
where
\begin{equation}
\Delta J^z = {(\pi Q)^{2}\over \omega}\sum_{k}k(|\tilde
A_{k}(\omega^{+}_k)|^{2}-|\tilde
A_{-k}(\omega^{+}_{-k})|^{2}),
\label{b8}
\end{equation}
\begin{equation}
\Delta J^{-1}={(\pi Q)^{2}\over
  \omega}\sum_{k}\sqrt{(2-k)(3+k)}(\tilde A_{k}(\omega^{+}_k)\tilde
A_{k+1}(\omega^{+}_k)^{*}+\tilde A_{-(k+1)}(\omega^{+}_{-k})\tilde
A_{-k}(\omega^{+}_{-k})^{*},
\label{b9}
\end{equation}
and $\Delta J^{1}=-(\Delta J^{-1})^{*}$. Note that to obtain the
latter relation we make a change $k\rightarrow -k$ in the summation
series (\ref{b5}) when considering the term proportional to ${\bf
  e}_{1}$. The Cartesian components of the vector of the angular
momentum transfer, $\Delta J^{x}$ and $\Delta J^{y}$, can be found from 
the relations $\Delta J^{x}={1\over \sqrt 2}(\Delta J^{-1}-\Delta
J^{1})$ and $\Delta J^{y}={i\over \sqrt 2}(\Delta J^{-1}+\Delta
J^{1})$. We have
$$ \Delta J^{x}={(\pi Q)^2\over \omega}\sum_k
\sqrt{(2-k)(3+k)}(\tilde R_k(\omega^+_k)\tilde R_{k+1}(\omega^+_k)+
\tilde I_k(\omega^+_k)\tilde I_{k+1}(\omega^+_k)
$$
\begin{equation}+\tilde
R_{-k}(\omega^+_{-k})\tilde R_{-(k+1)}(\omega^+_{-k})+
\tilde I_{-k}(\omega^+_{-k})\tilde I_{-(k+1)}(\omega^+_{-k})),
\label{b10}
\end{equation}
and 
$$ \Delta J^{y}={(\pi Q)^2\over \omega}\sum_k
\sqrt{(2-k)(3+k)}(\tilde I_{k+1}(\omega^+_k)\tilde R_k(\omega^+_k)
-\tilde I_k(\omega^+_k)\tilde R_{k+1}(\omega^+_k)$$
\begin{equation}
+\tilde I_{-(k+1)}(\omega^+_{-k})\tilde R_{-k}(\omega^+_{-k})-
\tilde I_{-k}(\omega^+_{-k})\tilde R_{-(k+1)}(\omega^+_{-k})),
\label{b11}
\end{equation}
where $\tilde R_{k}(\sigma)$ and $\tilde I_{k}(\sigma )$ are the real
and imaginary part of $\tilde A_{k}(\sigma )$, respectively, and we
remind that  $\tilde A_{k}(\sigma )$ is defined in equation (14).
Equations (\ref{b8}), (\ref{b10}) and (\ref{b11}) allows us to
determine all three components of the angular momentum transfer after 
a tidal encounter.   
 
\section*{Appendix C}
\section*{Energy transfer from a highly eccentric orbit associated with a normal mode
with general damping rate}
Here we consider the energy exchange associated with an orbit of high eccentricity
where the mode damping rate may be small. In this case the orbit is assumed to be quasi-periodic
and so  the system undergoes multiple close approaches in contrast to the single encounter
considered in parabolic limit that applies when the mode decay time is shorter
than the orbital period. The orbital period is assumed short enough so as to avoid the
stochastic regime (see Ivanov \& Papaloizou 2004). We obtain  an estimate for the factor
by which results for the parabolic case need to be multiplied in order to apply to this case.

We begin by writing down  equation (\ref{e13})
governing the response of a single mode amplitude allowing for 
mode damping (see Ivanov $\&$ Papaloizou 2004) in the form
\begin{equation}
\ddot b_n+\omega^2_0b_n+2(\gamma_n+i\omega^1_n)\dot b_n+2in\gamma_n\Omega b_n=S,  \quad  {\rm with }  \quad S=\int d^3x\rho
({\mbox{\boldmath $\xi$}}^{*}_n \cdot \nabla U), \label{AC1}
\end{equation}
where $\gamma_n$ is the mode damping rate.
Noting that the problem is very similar to that of a forced, damped
harmonic oscillator, we identify the mode energy as 
\begin{equation}
{\cal E}_n = \frac{1}{2}\left(|\dot b_n|^2+\omega^2_0|b_n|^2\right).  
\label{AC4}
\end{equation}
Its rate of dissipation through mode damping  follows from (\ref{AC1}) as
\begin{equation}
\frac{  d{\cal E}_n}{dt} =- 2\gamma_n |\dot b_n|^2-in\gamma_n\Omega (b_n(\dot b_n)^{*}-(b_n)^{*}\dot b_n),  
\label{AC41}
\end{equation}
see e.g. Ivanov $\&$ Papaloizou 2004.

We consider forcing in a periodic orbit with period $P_{orb}= 2\pi/\Omega_{orb}.$
The potential $U$ will then have a discrete Fourier series representation in the form
\begin{equation}
U = \sum_{k=-\infty}^{\infty} {\cal U}_k\exp{(-{\rm i}k\Omega_{orb}t}),  
\label{AC5}
\end{equation}
where the Fourier amplitudes are given by
\begin{equation}
 {\cal U}_k = \frac{\Omega_{orb}}{2\pi}\int^{\pi/\Omega_{orb}} _{-\pi/\Omega_{orb}} U \exp{({\rm i}k\Omega_{orb}t)}dt.
\label{AC6}
\end{equation}
Note that as the period tends to $\infty,$ or $\Omega_{orb} \rightarrow  0,$ with $k\Omega_{orb}$ remaining finite,
$2\pi{\cal U}_k/\Omega_{orb}$  approaches the continuous Fourier transform
associated with a parabolic orbit.

Because we consider an essentially linear response,  equations (\ref{AC1})-(\ref{AC4})
may be applied to each Fourier component separately and then the results combined.
Denoting by $S_k,$ the quantity $S$ evaluated for the corresponding term in the Fourier 
expansion for $U,$
and the associated   mode  response amplitude by $b_{nk},$  we have
\begin{equation}
 b_{nk} = \frac{S_k}{D}, \quad D=\omega_0^2-k^2\Omega_{orb}^2+2\omega^1_nk\Omega_{orb}-2i\gamma_n(k\Omega_{orb}-n\Omega).
 \label{AC7}
\end{equation}
As we discuss below only contributions in summation series over $k$, where $\omega_{0}\sim k\Omega_{orb}$ are 
important for our purposes. This means that the term proportional to $\gamma_n\Omega $ in the expression 
for $D$ in (\ref{AC7}) is much smaller than other terms as long as $\Omega \ll \omega_0$. It is neglected 
below. In the same approximation the term proportional to $\Omega $ in (\ref{AC41}) can be neglected as well
(in fact these terms could be retained in the analysis but little is gained).  

Note that in order to get (\ref{AC7}) we assume that the mode damping rate is approximately constant. This is not
formally the case when the dissipation is determined by non-linear processes considered in the text and $\gamma_n$ is
a function of mode amplitude.
 However, from equation (\ref{AC41}) it follows that the  characteristic time of  mode amplitude  decay 
is long compared with typical periods $1/(k\Omega_{orb})$ when $ \gamma_n /(k\Omega_{orb}) \ll 1$ 
which corresponds to the case of interest.
Then we can approximately 
consider $\gamma_n$ to be constant  and to correspond to the initial rate of
of mode decay,  $1/t_{nl},$ when dealing with  the Fourier harmonics with high value of $k$
that have frequencies corresponding to those of the excited normal modes.

We now evaluate the rate of energy dissipation by applying equation (\ref{AC4}) for each Fourier
component and summing over $k$ to give 
\begin{equation}
\frac{  d{\cal E}_n}{dt} =- 2\gamma_n \sum_{k=-\infty}^{\infty} \frac{ k^2\Omega_{orb}^2| S_k|^2}
{[[( k\Omega_{orb}- \omega^1_n-\omega_0 ) ( \omega^1_n-\omega_0-k\Omega_{orb} ) +(\omega^1_n)^2]^2+ 4\gamma_n^2k^2\Omega_{orb}^2] }.  
\label{AC8}
\end{equation}
We may make an estimate for the above sum by noting that for a  weakly damped  oscillation
mode,  most of the contribution to the sum is expected to come from values of $k$ corresponding to 
near resonance between a harmonic of the orbital period
and a mode oscillation period. Neglecting the inconsequential  quantity $(\omega^1_n)^2,$
which has been assumed to be negligible,
resonance  occurs when $k$ is such that  $k\Omega_{orb}= \omega^1_n \pm\omega_0,$
 which in fact corresponds to two modes, one with positive frequency and the other with
negative frequency.  As the important values of $k$ differ in sign, these can be dealt with separately
and the results summed. Without loss of generality, we shall restrict attention to the positive 
frequency mode for which values of $k$ near $\omega/\Omega_{orb}\gg 1$ are expected to dominate the sum
(we recall that $\omega =\omega_0 +\omega^1_n$) .
Assuming other quantities vary smoothly with $k,$ we replace $k$ by $k_0,$ being the closest integer to $\omega/\Omega_{orb},$
everywhere  except in the first bracket in the denominator of (\ref{AC8}) to obtain
\begin{equation}
\frac{  d{\cal E}_n}{dt} =- \frac{\gamma_n\omega^2| S_{k_0}|^2}{2\omega_0^2} \sum_{k=-\infty}^{\infty} 
\frac{1}{[( k\Omega_{orb}- \omega^1_n-\omega_0 )^2 +\gamma_n^2] }.  
\label{AC9}
\end{equation}
 The sum over $k$ is readily performed by standard complex variable techniques to give
 \begin{equation}
\frac{  d{\cal E}_n}{dt} =- \frac{\pi\omega^2| S_{k_0}|^2}{2\omega_0^2\Omega_{orb}}{\cal F} ,  
\label{AC10}
\end{equation}
where
 \begin{equation}
{\cal{ F}} = \frac{(\exp(4\pi\gamma_n/\Omega_{orb})-1)}{[\exp(4\pi\gamma_n/\Omega_{orb})+1 
-2\exp(2\pi\gamma_n/\Omega_{orb})\cos(2\pi\omega /\Omega_{orb}  )]} .  
\label{AC11}
\end{equation}
To interpret (\ref{AC10}) we remark first that when $2\pi \gamma_n/\Omega_{orb}$ is large corresponding
to  a damping time  shorter than the orbital period, ${\cal F}=1.$
Then we can integrate  (\ref{AC10}) over an orbital period  to obtain $\Delta ({\cal E}_n) \propto | S_{k_0}|^2/\Omega_{orb}^2,$
which ultimately  depends only on the continuous Fourier transform
of the perturbing potential. This limit  corresponds to a parabolic encounter.
More generally ${\cal F}$ gives an estimate of the factor by which the energy exchange
is modified when the damping time is  longer than an orbital period and the motion can be regarded a quasi periodic.
When $2\pi \gamma_n/\Omega_{orb}$ is small, we have 
 \begin{equation}
{\cal{ F}} \sim \frac{(\pi\gamma_n/\Omega_{orb})}{\sin^2(\pi\omega /\Omega_{orb} )} .  
\label{AC12}
\end{equation}
This factor could be large very  close to resonances where $\omega =k \Omega_{orb},$ for some integer $k.$
However, between resonances ${\cal F}$ has a minimum value $\gamma_n P_{orb}/2,$
while a reciprocal mean over mode frequencies would suggest a value  $\gamma_nP_{orb}.$
As orbital evolution very close to resonances is expected to be rapid, the latter values are expected to be
more appropriate for estimating evolution times.   Recall that for comparison with section \ref{KGS}, 
the mode decay rate $\gamma_n$  adopted here here should be identified with $1/ t_{nl}$ used  there.

\end{appendix}

\end{document}